\documentclass[10pt,conference,anonymous]{IEEEtran}
\IEEEoverridecommandlockouts
\usepackage{cite}
\usepackage{amsmath,amssymb,amsfonts}

\usepackage{algorithmic}

\usepackage{pgfplots}
\usepackage{tikz}
\usepackage{arydshln}

\usepackage{graphicx}
\usepackage{textcomp}
\usepackage{xcolor}
\def\BibTeX{{\rm B\kern-.05em{\sc i\kern-.025em b}\kern-.08em
    T\kern-.1667em\lower.7ex\hbox{E}\kern-.125emX}}

\usepackage[ruled,vlined,linesnumbered]{algorithm2e}
\usepackage{amsthm}
\theoremstyle{plain}

\newtheorem*{problem*}{Problem}

\usepackage{enumitem}

\usepackage{epstopdf}
\usepackage{multirow}
\usepackage{color,soul}
\usepackage{colortbl}
\usepackage{array} 
\graphicspath{{../pdf/}{../jpeg/}}
\DeclareGraphicsExtensions{.pdf,.jpeg,.png}
\usepackage{subfig}
\usepackage{float}
\usepackage{lscape}
\usepackage{tabularx}
\usepackage{tabulary}
\usepackage{footmisc}
\usepackage{balance}
\usepackage{booktabs}
\usepackage{longtable}
\usepackage{hyperref}

\newcommand{\approach}{DeepEST}

\newcommand{\STAB}[1]{\begin{tabular}{@{}c@{}}#1\end{tabular}}

\begin{document}

\title{Operation is the hardest teacher: estimating\\DNN accuracy looking for mispredictions\thanks{This work has been supported by the project COSMIC of UNINA DIETI.}}

\author{\IEEEauthorblockN{Antonio Guerriero, Roberto Pietrantuono, Stefano Russo}
\IEEEauthorblockA{\textit{DIETI, Universit\`a degli Studi di Napoli Federico II}\\
Via Claudio 21, 80125 - Napoli, Italy \\
\{antonio.guerriero, roberto.pietrantuono, stefano.russo\}@unina.it}
\vspace{-12pt}
}
\maketitle

\begin{abstract}
Deep Neural Networks (DNN) are typically tested for accuracy relying on a set of unlabelled real world data (operational dataset), from which a subset is selected, manually labelled and used as test suite. This subset is required to be small (due to manual labelling cost) yet to faithfully represent the operational context, 
with the resulting test suite containing roughly the same proportion of examples causing misprediction (i.e., failing test cases) as the operational dataset. 

However, while testing to estimate accuracy, it is desirable 
 to also learn as much as possible from the failing tests in the operational dataset, since they inform about possible bugs of the DNN.  
A smart sampling strategy may allow to intentionally include in the test suite many examples causing misprediction, thus providing this way more valuable inputs for DNN improvement while preserving the ability to get trustworthy unbiased estimates.  

This paper presents a test selection technique (\approach{}) that actively looks for failing test cases in the operational dataset of a DNN, with the goal of assessing the DNN expected accuracy 
 by a small and ``informative'' test suite (namely with a high number of mispredictions) for subsequent DNN improvement. 
Experiments with five subjects, combining four DNN models and three datasets, are described. The results show that \approach{} provides DNN accuracy estimates with precision close to (and often better than) those of existing sampling-based DNN testing techniques, 
while detecting from 5 to 30 times more mispredictions, with the same test suite size. 
\end{abstract}


\section{Introduction}
\label{Introduction}
\makeatletter{}
Deep Neural Networks (DNN) are today integral part of many applications, including safety-critical software systems such as in the medical \cite{Obermeyer2016} and autonomous driving domains \cite{Bojarski}. 
Testing is 
 a crucial activity in the development of such %
systems, for both quality/safety reasons to avoid %
DNN-caused catastrophic failures \cite{Amir2018,Stewart2018},  
and for cost as well --  very large
 samples may be needed to reliably test a DNN \cite{DeepTest,Taigman2014}.

A significant research effort is currently being put on DNN testing. 
A primary goal is to find adversarial examples causing \textit{mispredictions}, namely to expose as many failing behaviours as possible \cite{Pei19,Ma18,Zhang18,Ma2018combinatorial,Odena18}.
Several structural coverage criteria have been proposed to  drive the automated generation of test inputs and assess the test adequacy -- 
neuron coverage \cite{Pei19}, k-multisection neuron coverage, neuron boundary coverage \cite{Ma18}, combinatorial coverage  \cite{Ma2018combinatorial}.  %
It has been argued that these criteria may be misleading,  %
because of the low correlation between the number of misclassified inputs in a test set and its coverage \cite{Li19Structural}. 
Wu \textit{et al.} \cite{Wu2019} and Kim \textit{et al.} \cite{Kim19} recently considered discrepancy measures between the training/validation data and test data, to improve fault detection and to have coverage criteria better correlated to failure-inducing inputs. 

The output of this type of failure-finding testing (and then debugging) process is an improved DNN, %
with higher accuracy. 
This resembles what is called \textit{debug testing} in the traditional testing literature \cite{Frankl1998}. %
Beside differences in testing DNN-based and conventional software (e.g., the oracle definition), which make DNN testing problematic, a further issue is that 
the so-obtained testing results  
 are not necessarily related to the accuracy actually experienced in operation. In fact, \textit{testing data may be not representative of the actual operational context}. This may happen when test data are generated artificially (like in adversarial examples generation) 
or they differ significantly from input observed in the field.  
   The number of %
 mispredictions and/or the 
 coverage achieved give only an ``indirect'' (and, for what said, inaccurate) measure of the expected accuracy in operation, and ultimately of the confidence that can be placed in the DNN. %

A less investigated research path is testing a DNN with the explicit goal of providing a statistical estimate of its \textit{expected accuracy} in the operational context. 
In software reliability engineering, this is a well established practice known as \textit{operational testing} \cite{Lyu1996}. 
The objective is to assess how well a DNN will perform in the intended context using a small yet effective amount of test data. 
With a cost-effective accuracy estimate, testers can establish a threshold-based release criterion and correct or tune their artificial networks (e.g., adjust the DNN structure or hyper-parameters) until the criterion is met. %
The reference scenario is the following: a DNN model is trained with a set of data (\textit{training dataset}) and is meant to operate in a given context (\textit{operational context}). 
To test the model, an arbitrarily large set of operational data is available (\textit{operational dataset}) containing examples whose correct label is unknown: the goal of the tester is to select a small subset of the operational data, to be manually labelled and used as test cases to estimate the accuracy of the model in operation. Due to manual labelling, testers typically look for a trade-off between a good estimate 
 and the labelling cost.  
This problem has been recently faced by Li \textit{et al.} \cite{Li19}. Their selection criterion is to minimise cross-entropy between selected test data and the whole set of unlabelled operational data, so as to  have a sample of tests 
representative of the operational context. 
As this approach does not explicitly look for failing examples, it does not give much room for improving the DNN accuracy. %
 From this viewpoint, many tests selected are useless, as they are non-failing examples that serve only the purpose of having highly-confident estimates. %

Given the above, we either have a test suite exposing failures but not representative (adversarial-like testing), or representative tests but few failures exposed (operational testing). 
This paper targets the drawbacks of both strategies together.  
We present \approach{} (Deep neural networks Enhanced Sampler for operational Testing), 
an operational testing method for DNN that builds test suites from an operational dataset so as to provide a close and efficient estimate of the expected accuracy \textit{and} to find, at the same time, a high number of failing examples. %
Paraphrasing a famous aphorism by O. Wilde 
 (``Experience is the hardest kind of teacher. It gives you the \textit{test} first and the \textit{lesson} afterward."), \approach{} aims to sample, from operational data, many failing \text{tests} to learn from %
 while building a set of tests suited for the DNN accuracy estimate.  

With respect to pure operational testing, \approach{} is expected to provide DNN accuracy estimates that are more precise and more efficient (for number of tests required), plus the practical advantage of enabling debug-testing-like scenarios (improving accuracy through debugging/re-training). 
With respect to adversarial-like testing, \approach{} is able to provide 
accuracy estimates, enabling evaluation of alternative designs or DNN fine tuning, and establishing a release criterion directly related to what will be observed in operation. 

Experiments with various DNN and datasets are presented, which assess \approach{} effectiveness, sensitivity to the number of tests to select 
from the operational dataset, and dependency on the dataset. 
As \approach{} 
can exploit various types of 
auxiliary information to select tests, the experimental study considers four variants. %
The results show that all the variants
produce accuracy estimates similar to those of the state-of-the-art technique, 
while detecting from 5 to 30 times more mispredictions, with the same sample size. 

The paper is structured as follows. Section \ref{Background} introdcues sampling-based operational testing concepts used by \approach. 
Section \ref{Method} describes the \approach{} algorithm. Sections \ref{Evaluation} and \ref{Results} report the experimentation. %
Section \ref{Related} discusses related work. 
Section \ref{Conclusion} concludes the paper.

\section{Sampling-based operational testing of DNNs}
\label{Background}
\makeatletter{}
\subsection{Operational testing of DNNs}
In the traditional testing literature, \textit{operational testing} refers to the family of techniques that use an operational profile to test a system to estimate %
 its expected \textit{reliability} (i.e., probability of not failing) in operation.  
Likewise, the primary goal of \textit{DNN operational testing} is to estimate the expected \textit{accuracy} %
(i.e., probability of not having mispredictions) in a given operational context \cite{Li19}.\footnote{We use the terms misprediction and failure interchangeably for DNN.} 
Two main challenges arise. 

\noindent \textbf{Data skew}. The idea of operational testing is that testing should not just care about exposing possible failures, but should be able to spot those failure-causing inputs that are more likely to occur in operation. 
In case of a relevant mismatch between the pre-release test data and the post-release context, the system could be stimulated in operation with inputs never seen during testing, with unexpected failures. 
 
Data skew is a concern for DNN, more than for traditional software systems. These are expected to work on a range of input data given to functionalities, and can be tested on a small carefully-selected sample of input data (e.g., obtained by input space partitioning). 
DNN are data-driven by nature; they are constructed around a training dataset, %
and generalising beyond what observed during training is hard. This data-driven approach is governed by a statistical process and, due also of the black-box nature of DNN, it is tricky to identify classes of inputs that homogeneously represent the expected behaviour in operation. For instance, white-box partitioning 
 has been shown to be not clearly correlated to the failing behaviour \cite{Li19Structural}. 
Thus, a \textit{drift} of the post-release operational context from the pre-release testing context is more likely to cause unexpected failures compared to traditional software.

\noindent \textbf{The \textit{imitation bias} of operational testing.} 
The operational context drift is just a triggering condition for unexpected failures at runtime. The problem occurs because of the way in which operational testing is conducted. Operational testing selects a small data sample that can accurately represent the population; however, the mere imitation of the expected input can be inefficient, especially in highly reliable systems, because many failure-free tests are executed to get an acceptable estimate. 
A representative sample would roughly contain the same proportion of examples causing misprediction as the operational dataset. 
There are two problems with this: 
first, in highly accurate DNNs, the number of examples causing mispredictions is low, thus requiring other types of testing activities dedicated to detect mispredictions (e.g., through adversarial-like techniques) to possibly improve the accuracy.
 Second, just mimicking the expected usage is fine from the estimation point of view \textit{as long as} the imitation is faithful. If this is not the case, the risk of overestimation increases: 
if we only aim at having a representative sample of tests, 
 the actual experienced accuracy may be significantly smaller than the estimated one if the operational context drifts, because of the occurrence of unexpected mispredictions. %

Thus, a conventional approach for DNN operational testing allows to obtain the desired accuracy estimate but, since it reveals few mispredictions, can be ineffective (leading to wrong estimates) 
and inefficient (requiring further separate testing activities to detect mispredictions). %
Recent results in operational testing for traditional software show that exposing failures and estimating reliability are not contrasting objectives \cite{TSE2016}. %
A strategy 
 that actively looks for failures (rather than just mimicking the expected usage) can lead to accurate and stable estimates and, at the same time, expose many failures. %
Our aim is exactly to spot failing examples in a DNN operational dataset, while preserving the ability to yield effective (small error) and efficient (low variance)  estimates. 

\subsection{Sampling-based testing}
\approach{} uses statistical sampling, %
a natural way to cope with estimation problems: it serves to design sampling plans tailored for a population under study, providing effective and efficient estimators. 
In sampling-based testing  \cite{ISSRE2016}, the sample is the set of $n$ test cases $T$=$\{t_1, \dots, t_n\}$, having a binary outcome (pass/fail). Test outcomes are a series of independent Bernoulli random variables $z_{t_i}$  such that $z_{t_i}$=$1$ if the DNN predicts the correct label for $t_i$, $z_{t_i}$=$0$ otherwise. 
The parameter of our interest is the DNN accuracy $\theta$;  we aim to an estimate $\hat{\theta}$ with two desirable properties: \textit{unbiasedness} -- i.e., the expectation of the estimate $\mathbb{E}[\hat{\theta}]$ should be equal to the true value $\theta$ - and \textit{efficiency} -- for the given the sample size, the variance of $\hat{\theta}$ should be as low as possible (for a highly confident, stable estimate). 
The probability that $z_{t_i}=1$ corresponds to the true (unknown) proportion: $\theta = \frac{\sum_{t=1}^{N}{z_{t_i}}}{N}$, with $N$ being the population size (i.e., the size of the operational dataset). 
 
Simple random sampling with replacement (SRSWR) is the baseline approach: an unbiased  estimator of $\theta$ is the observed proportion of correct predictions over the number of trials $n$: 
 
\begin{equation}
\hat{\theta}_{SRSWR} = \frac{\sum_{t=1}^{n}{z_{t_i}}}{n}. 
\label{thetaAssessment}
\end{equation}
Having assumed independent variables, the variance of  $\hat{\theta}$ is:
\begin{equation}
V(\hat{\theta}_{SRSWR}) =   \frac{\theta( 1- \theta)}{n}.
\label{VarAssessmentSRSWR}
\end{equation}
An improvement is represented by simple random sampling \textit{without replacement} (SRSWOR), namely, the same test case is not selected twice: this reduces the variance to: 
\begin{equation} 
\begin{array}{l}
V(\hat{\theta}_{SRSWOR})  = \frac{N - n}{N -1} \frac{\theta( 1- \theta)}{n}.
 \end{array}
\label{VarAssessmentSRSWOR}
\end{equation}
While SRS keeps the mathematical treatment simple, it is unable to exploit additional information a tester might have. 

Exploiting \textit{auxiliary information} to modify the sampling scheme is what is done in sampling theory to get more efficient estimators \cite{BOOK1}. The sampling is made proportionally to an auxiliary observable variable assumed to be related to the (unknown) quantity to estimate;  the estimator is then adjusted to account for the non-uniform sampling, so as to preserve unbiasedness. For instance, stratified sampling is a strategy which uses knowledge about which sample units are expected to have homogeneous values, and selects units contributing more to lower the estimate's variance. 

Li \textit{et al.} \cite{Li19} present two sampling strategies for DNN, which are, to our knowledge, the only attempt to DNN operational testing: Confidence-based Stratified Sampling (CSS) and Cross Entropy-based Sampling (CES) -- the latter being the authors' proposal. Both strategies exploit auxiliary information to drive the sampling task. 

In CSS, sampling  is proportional to the confidence value provided by classifiers when predicting a label: examples with higher confidence are more likely to be selected as part of the test suite. This works well when the classifier is reliable, namely if examples with higher confidence are actually those for which the prediction is more likely to be correct (in other words, the model is perfectly trained for the operation context). Whenever the operation context drifts from the training one, CSS exhibits poor performance. 

CES attempts to overcome this limitation by using the output of the $m$ neurons in the last hidden layer in the last hidden layer, assumed to be more robust to the operation context drift.  It builds the test suite trying to minimize the average cross-entropy between the probability distribution of the $m$-dimensional representation of the output of neurons computed on the operational dataset and on the selected tests. 

To pursue the double objective of sampling cases causing mispredictions and estimating accuracy efficiently, \approach{} adopts an \textit{adaptive} and \textit{without-replacement} sampling algorithm, described in Section \ref{Method}.

\subsection{Auxiliary information}
To look for mispredictions, the auxiliary information leveraged by \approach{} adaptive sampling should represent the belief about some factor(s) related to the model's (in)accuracy. %
We consider two auxiliary variables, related to somehow opposite sources of information: the \textit{confidence} value
 and the \textit{distance} between the operational dataset example and the training dataset. The latter is based on the result by Kim \textit{et al.} \cite{Kim19} that inputs more ``distant'' from training data are more likely to cause misprediction -- they show that distance is correlated to mispredictions, which is what we look for. We borrow their distance metrics \textit{Distance-based Surprise Adequacy} (DSA) and \textit{Likelihood-based Surprise Adequacy} (LSA).
These are computed by the \textit{Activation Trace} (AT), namely a vector of activation values of each neuron of a certain layer corresponding to a certain input; we compute AT with reference to the last DNN activation layer. 
LSA is defined as the negative log of density (computed via Kernel Density Estimation). DSA is defined as the Euclidean distance between the AT of a new input and ATs observed during training.
The variant using confidence is \approach{}$_{CS}$; the variants using the distance metrics are \approach{}$_{DSA}$ and \approach{}$_{LSA}$. 

Performance of an auxiliary variable can strongly depend on the DNN and on the training/operation dataset: for instance, for a distance metric the belief that examples far from the one in the training set are more likely to cause a misprediction is not an absolute truth. 
In particular, if an example is very similar to many others in the training set according to the distance metric, but has a different label, distance will be not a good metric to select it. 
Similarly, the confidence is a good proxy \textit{if} the DNN is well trained for the operational context. 
In general, relying on a single auxiliary variable may work well in some settings and bad in others. Combining multiple beliefs is a choice that is expected to improve the stability of results across multiple settings. 
Based on this, we define a further variant of our algorithm, named \approach{}$_C$, considering as auxiliary variable the combination of confidence and distance: 
\begin{equation}
	P = P_c \times (1-P_d)
\end{equation}
where $P_c$ is the confidence value, $P_d$ is the DSA value normalized [0, 1]\footnote{In \approach$_C$, DSA is preferred to LSA since it has been shown to have better performance  for the deeper layer \cite{Kim19}.} and $P$ is probability of correct prediction. 
The intuition behind is that the probability of a correct prediction is related to both the confidence of the DNN and the \textit{drift} of the example from what seen during training. In fact, $P_c$ is related to the probability that the DNN does a correct prediction \textit{according to what seen during training} (in other words: it is the probability of correct prediction with perfect training); $P_d$ is a proxy for the probability of  wrong prediction related to how far the example is from what seen during training -- hence due to the \textit{imperfection} of training. If confidence $P_c$ is high \textit{and} the example is close to the training dataset (i.e., $P_d$ is small), there is a high chance of correct prediction.

\section{The \approach{} algorithm}
\label{Method}
\makeatletter{}

Both CES and CSS select a sample of tests representative of the operational context. \approach{} takes a different approach, favouring the sampling of failing (namely, \textit{misprediction-causing}) tests, then used to 
 estimate the DNN accuracy. 
While the estimation ability demands for probabilistic (hence ``sampling-based'') selection of examples, the very idea of \approach{} is that, 
since mispredictions are usually rare compared to correct examples, looking for failing tests is well handled by \textit{adaptive sampling} \cite{BOOKThompson}: the examples progressively selected may give hints about the probability of finding other failing examples, so as to adapt the sampling process accordingly. Given a samples size, adaptive sampling implicitly assumes that inputs of interest are not uniformly spread across the input space, and adopts a disproportional selection to spot them, counterbalanced by the estimator to preserve unbiasedness. In software testing, this is expected to give smaller variance compared to conventional sampling (e.g., SRS), since the failing inputs are usually not uniformly distributed.  
The \approach{} sampling algorithm is inspired to the Adaptive Web Sampling \cite{Thompson}, a flexible design for sampling rare populations. 

The above auxiliary variables are used to define a \textit{weight} $w_{i,j}$ between any pair of examples $i$ and $j$ of the operational dataset, 
used to explore the example space adaptively. 
For instance, let us assume to use the normalised DSA \textit{distance}: if the example $i$ has distance $P_{d_i}$ (representing the belief that $i$ causes a misprediction), and a pre-defined threshold $\tau$ is exceeded (i.e., $i$ has a sufficiently high distance compared to the others), then all the $w_{i,j}$ values ($\forall j$ of the operational dataset) are set to their distance $P_{d_j}$; otherwise $w_{i,j}=0$. 
This way, a strong-enough belief about example $i$ causing misprediction entails the activation of all the weights toward $i$. The latter are used, as explained hereafter, for sampling, and makes the algorithm 
follow the distance criterion to spot potential clusters of failing examples. 

The \approach{} sampl\-ing strategy is sketched in Algorithm 1. 
Assuming $n$ 
examples 
to be selected from the operational dataset as test cases, the algorithm selects and executes one test case per step. The first input is selected by simple random sampling, namely, initially all examples have equal probability of being selected. 
Then, one of two sampling schemes is used to select next test: \textit{weight-based sampling} (WBS), or \textit{simple random sampling} (SRS). %
Example $i$ is selected from the operational dataset at step $k$ with probability $q_{k,i}$ given by: 
\begin{equation}
q_{k,i}=r \cdot \frac{\sum_{j \in s_k} w_{i,j}}{\sum_{h \notin s_k, j \in s_k} w_{h,j}}+(1-r) \cdot \frac{1}{N-n_{s_k}} 
\label{eq1}
\end{equation}
\noindent where:
\begin{itemize}
	\item $r$: probability of using WBS (hence, probability of using SRS = $1$-$r$; $r$ is set to 0.8 in our implementation);
	\item $s_k$: current sample (all examples selected up to step $k$); 
	\item $w_{i,j}$: weight relating example $j$ in $s_k$ to example $i$; 
	\item $n_{s_k}$: size of the current sample $s_k$;
	\item $N$: size of the operational dataset. 
\end{itemize}
For both WBS and SRS, the selection is \textit{without replacement}.\footnote{Without replacement sampling schemes generally give smaller variance than their with replacement counterpart on the same sample size \cite{BOOK1}.}
WBS selects an example $i$ proportionally to the sum of weights $w_{i,j}$ of already selected examples toward $i$ -- the chance of taking $i$ depends on the current sample, favouring the identification of clusters of failing examples \textit{if} the auxiliary variable (hence, $w_{i,j}$) is well-correlated with mispredictions.  
As this is not always the case,  the WBS ``depth'' exploration is balanced by SRS, chosen with probability ($1$-$r$), for a breadth exploration of the example space. This diversification in the search is useful to escape from unproductive cluster searches.  The steps are repeated until the testing budget $n \le N$ is over.

\begin{algorithm}[t]
\caption{\textbf{\approach{} adaptive sampling}}
\label{alg:Algorithm}
\KwData{$OD$: operational dataset; $i$: current sample; $T$: test suite; $n$: number of tests; $r$: probability of  WBS}
$T=\emptyset$\;
$i=$ \textit{SRS\_sampling(OD)};\tcp{first sample by SRS} 
$OD=OD\setminus i$;\tcp{remove sample from dataset}
$T=T\cup i$; \tcp{add to suite}
$y_1 =$ \textit{labelling\_and\_test\_execution(i)}\;

\For{$k=2; k <= n;$ k++}{
	$rs = random(0, 1)$\;
	\eIf{$rs < r$}{
		$i = $ \textit{WBS\_sampling(OD);}  $OD= OD\setminus i$; %
	}{
		$i = $ \textit{SRS\_sampling(OD);}   $OD = OD\setminus i$; %
	}
	$T=T\cup i$\;
	$y_i = $ \textit{labelling\_and\_test\_execution(i)}\;
	$z_k = $\textit{ Equation} \ref{zk}\;
}
$\hat{\theta} =$ \textit{Equation} \ref{estimation}; \tcp{compute final estimate}
\end{algorithm}

At step $1$,
the %
 probability that a randomly selected example will cause a misprediction is estimated as the outcome $y_1$ of the first test (1 in case of misprediction, 0 otherwise). 

At step $k$$>$$1$, example $i$, whose outcome is $y_i$, is selected with probability $q_{k,i}$ according to Eq. \ref{eq1}, and the estimator of the misprediction probability is that by Hansen-Hurwitz \cite{Hansen1943}: 
\begin{equation}
\label{zk}
z_k= \frac{1}{N}(\sum_{j \in s_{k}} y_j + \frac{y_i}{q_{k,i}})
\end{equation}
\noindent where %
the $y_{j}$ values are the outcome of the tests already selected.  %
$z_k$ is an unbiased estimator of the expected misprediction probability at step $k$; 
the final estimator of the expected \textit{accuracy} of the DNN is 1 minus the average of the $z_k$ values: 
\begin{equation}
\label{estimation}
\hat{\theta} = 1 - \frac{1}{n}(y_1  + \sum_{k=2}^{n} z_{k}) 
\end{equation}
\noindent where $n$ is the number of tests run.

\section{Evaluation}
\label{Evaluation}
\makeatletter{}
\subsection{Experimental subjects} 
The four variants of \approach{} are evaluated against the SRSWR scheme as baseline and the mentioned state-of-the-art technique CES. 
Five experiments are conducted with four DNN models and three popular datasets. 

The datasets are MNIST, a dataset of handwritten digits \cite{Lecun98}; 
CIFAR10, for image processing systems; 
and CIFAR100, similar to the previous one but with 100 classes \cite{Krizhevsky09}.
The chosen DNN models are ConvNet5 (here simply CN5) %
and LeNet5 %
for MNIST classification; ConvNet12 (simply, CN12) %
and VGG16 for CIFAR10 classification; VGG16 for CIFAR100.\footnote{CN5 and CN12 are calibrated in the same way as Li \textit{et al.} \cite{Li19}; LeNet5 is calibrated as Kim \textit{et al.} \cite{Kim19}; for the VGG16 network we considered the weights at \url{https://github.com/geifmany/cifar-vgg}.}
Table \ref{datasets} lists the five subjects (DNN-dataset pairs); the true accuracy is in the last column. 

\begin{table}[t]
	\centering
	\caption{Experimental DNN and datasets}
	\label{datasets}
	\def\arraystretch{1.2}%
	\begin{tabular}{@{}l|c|c|c|c|c|c@{}}
\toprule
	\multicolumn{3}{c|}{\textbf{Subject}}              & \textbf{}              & \textbf{Training} & \textbf{Test}       & \textbf{True}\\ 
	\textbf{} & \textbf{DNN} & 	\textbf{Dataset} & \textbf{Classes} & \textbf{set size}  & \textbf{set size} &  \textbf{accuracy}\\ \hline
	{S1} & {CN5} & \multirow{2}{*}{MNIST} &\multirow{2}{*}{10} &\multirow{2}{*}{60,000} &\multirow{2}{*}{10,000}& 0.9905 \\ \cline{1-2}\cline{7-7} 
	{S2} & {LeNet5} &&&&& 0.9868 \\ \hline
	{S3} & {CN12} & \multirow{2}{*}{CIFAR10} &\multirow{2}{*}{10}& \multirow{2}{*}{50,000}& \multirow{2}{*}{10,000}& 0.8066 \\ \cline{1-2}\cline{7-7} 
	{S4} & {VGG16} & &&&& 0.9359 \\ \hline
	{S5} & {VGG16} & {CIFAR100}   &100 & 50,000 & 10,000& 0.7048 \\\bottomrule
	\end{tabular}
\end{table}

\subsection{Research questions and experiment design}
The 
 evaluation 
  answers the following research questions. 

\vspace{3pt}
\noindent \textbf{RQ1: Effectiveness}. \textit{How does \approach{} perform in finding inputs causing misprediction (i.e., failing examples) and simultaneously estimating a DNN operational accuracy?}

\vspace{3pt}
To gauge \approach{} ability to %
provide effective DNN accuracy estimates with few examples, while spotting a high number of failing inputs, 
we set to 200 the number of tests to select (then varied to answer RQ2) and repeat 30 times the execution of the 6 compared techniques on the 5 subjects. 

As for evaluation metrics, we compute: 
\begin{itemize}  
\item The accuracy $\hat{\theta_i}$ at the $i$-th repetition, and then compute the %
 Mean Squared Error (MSE) as $MSE(\hat{\theta}) = \frac{1}{30} \sum_{i=1}^{30} (\hat{\theta}_i -\theta)^2$, where $\theta$ is the true operational accuracy. 
Note that for unbiased estimators, MSE and variance can be considered indistinguishable. In fact, $MSE$ = $Variance$+$Bias^2$ and $Bias(\hat{\theta})=\mathbb{E}[\hat{\theta}] - \theta=0$. 

The precision of the estimator is: $\pi(\hat{\theta})$ = $\frac{1}{MSE(\hat{\theta})}$, and the relative precision (or relative efficiency) of estimator $A$ with respect to $B$ is:  $\pi_{A,B}$ = $\frac{MSE(\hat{\theta}_B)}{MSE(\hat{\theta}_A)}$ ($\pi_{A,B}>1$ means that $A$ is better than $B$). 
\item The average 
 number of failures ($\varphi$ = $Mean(\varphi_i)$) 
  with $\varphi_i$ being the number of failures in repetition $i$.  %

For comparison purpose, we consider the relative number of failures of technique $A$ with respect to $B$: $\rho_{A,B} = \frac{\varphi_A}{\varphi_B}$ ($\rho_{A,B}>1$ means that $A$ is better than $B$). 
 \end{itemize}

\vspace{6pt}
\noindent \textbf{RQ2: Sensitivity to sample size}.  
\textit{How does the performance of \approach{} vary with the sample size?}

\vspace{3pt}
It is important to figure out how performance varies with the number of test cases to select from the operational dataset, namely with sample size. Indeed, \approach{} aims to perform well especially with a small sample size, so as to yield precise estimates with relatively few examples to be manually labelled.

\vspace{6pt}
\noindent \textbf{RQ3: Dataset influence}.  \textit{How is \approach{} performance affected by the datasets?}

\vspace{3pt}
In DNN testing, results are often heavily dependent on the (training and operational) datasets. 
This RQ aims to figure out how these may influence the ability of the auxiliary variables (confidence, DSA, LSA) to discriminate failing examples, affecting the performance of \approach{}. 
To answer RQ3, the test %
set is completely labeled, so as to identify all failures. 

\subsection{Implementation}
\approach{} is implemented mostly in Java.\footnote{A replication package is at: \url{https://github.com/dessertlab/DeepEST}}
The implementation of the distance metric in \approach{}$_{DSA}$ and \approach{}$_{LSA}$ is the same used by Kim \textit{et al.} \cite{Kim19}; we used their Python scripts to compute DSA and LSA values. 
These are computed considering the last \textit{activation layer} of each DNN.
The threshold $\tau$ needed for the weights definition is set as follows: 
\begin{itemize}
	\item \approach{}$_{DSA}$: $\tau = $ $mean(DSA)+2\times Std(DSA)$; %
	\item \approach{}$_{LSA}$:  $\tau = $ $mean(LSA)+Var(LSA)$.
\end{itemize}
The threshold for \textit{confidence}, used by \approach{}$_{CS}$ and \approach{}$_{C}$,  is set to $0.7$, assuming that lower confidence values are more related to misprediction (i.e., the weights are activated when the confidence is less than $\tau=0.7$). %

In CES, the selection exploits the output of the last hidden layer. 
We use the same configuration as the original article \cite{Li19}. 
The size of the initial sample is $p$=$30$, enlarged by a group $Q^*$ of $q$=$5$ examples at each step. The number of random groups from which $Q^*$ is selected is $L=300$. 
For CES and SRS, we used the Python scripts provided by Li et al. \cite{Li19}.

\section{Results}
\label{Results}
\makeatletter{}

\begin{figure*}[t]
	\centering
	\subfloat[Subject S1 (CN5, MNIST)]{\includegraphics[width=0.3\textwidth]{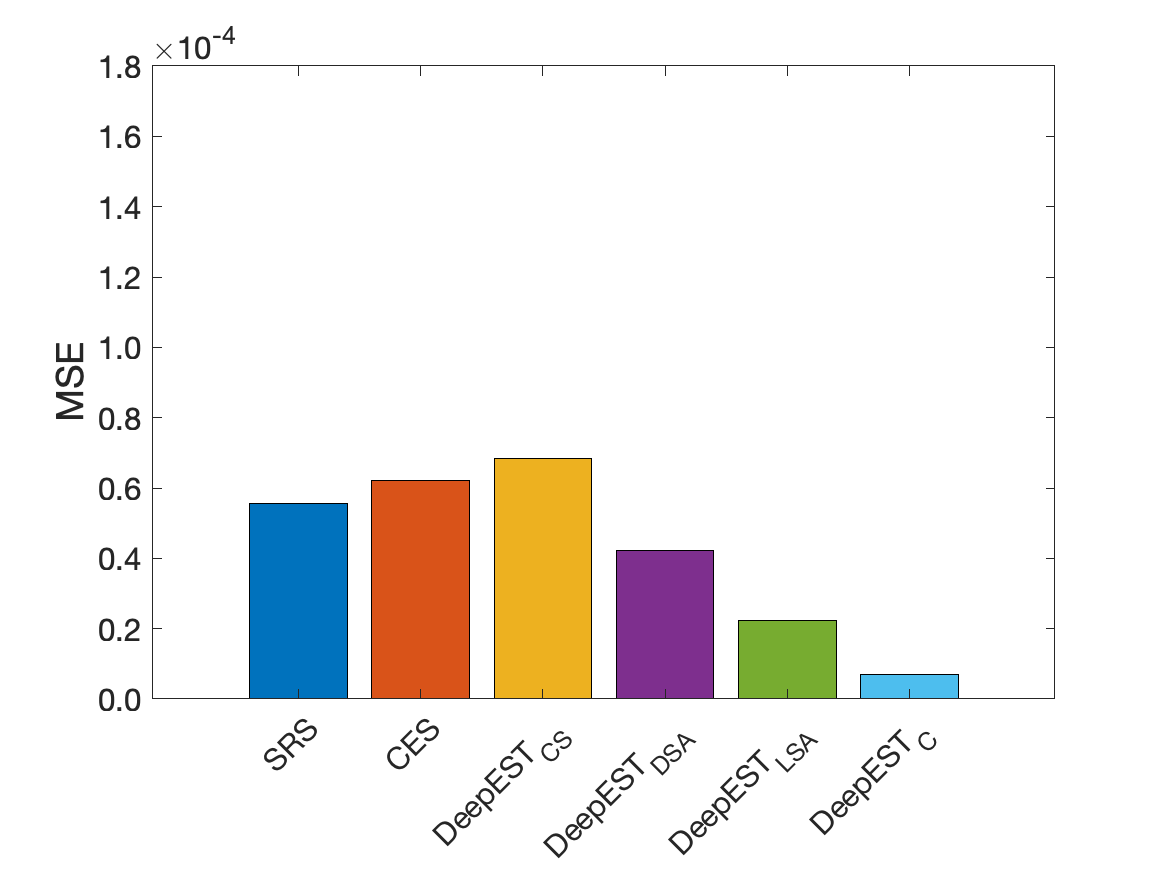}
		\label{1MSE}\hspace{-0,5cm}}
	\subfloat[Subject S2 (LeNet5, MNIST)]{\includegraphics[width=0.3\textwidth]{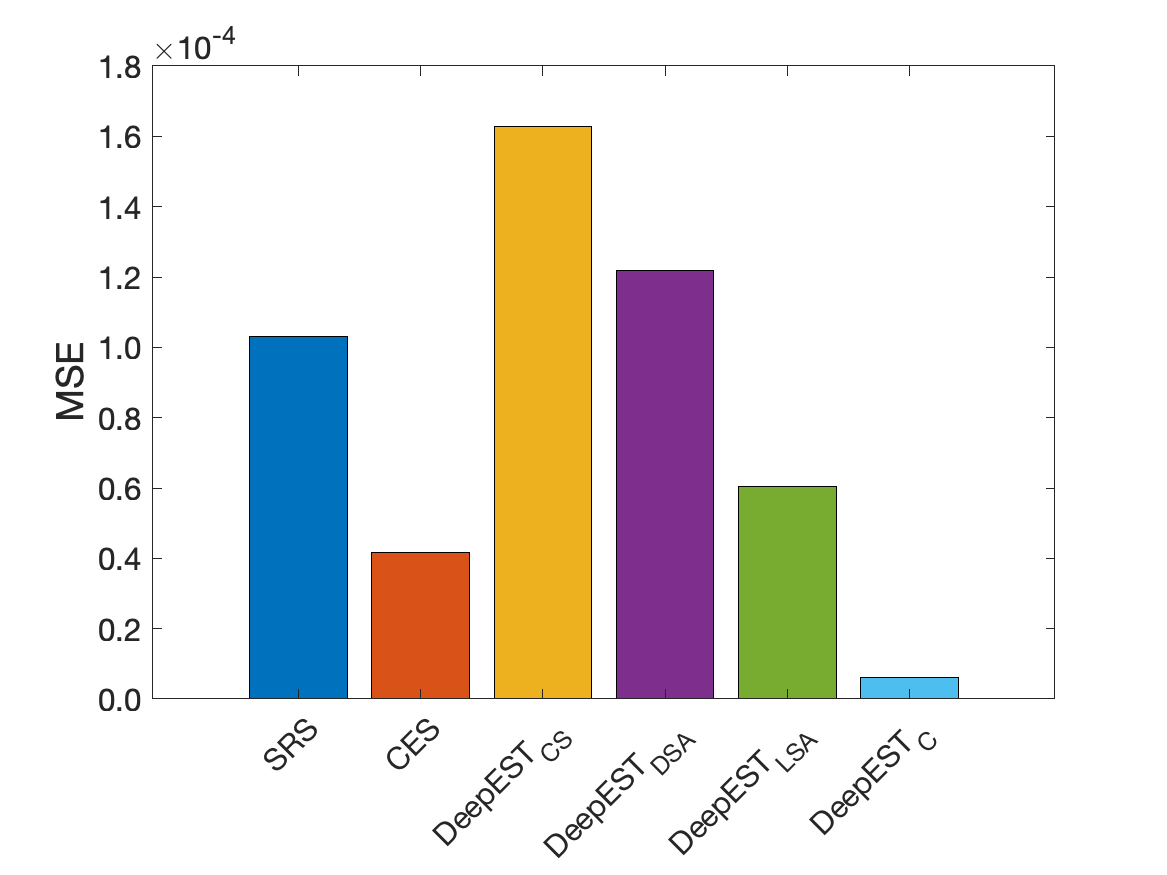}
		\label{2MSE}}\\
\vspace{-8pt}
	\subfloat[Subject S3 (CN12, CIFAR10)]{\includegraphics[width=0.3\textwidth]{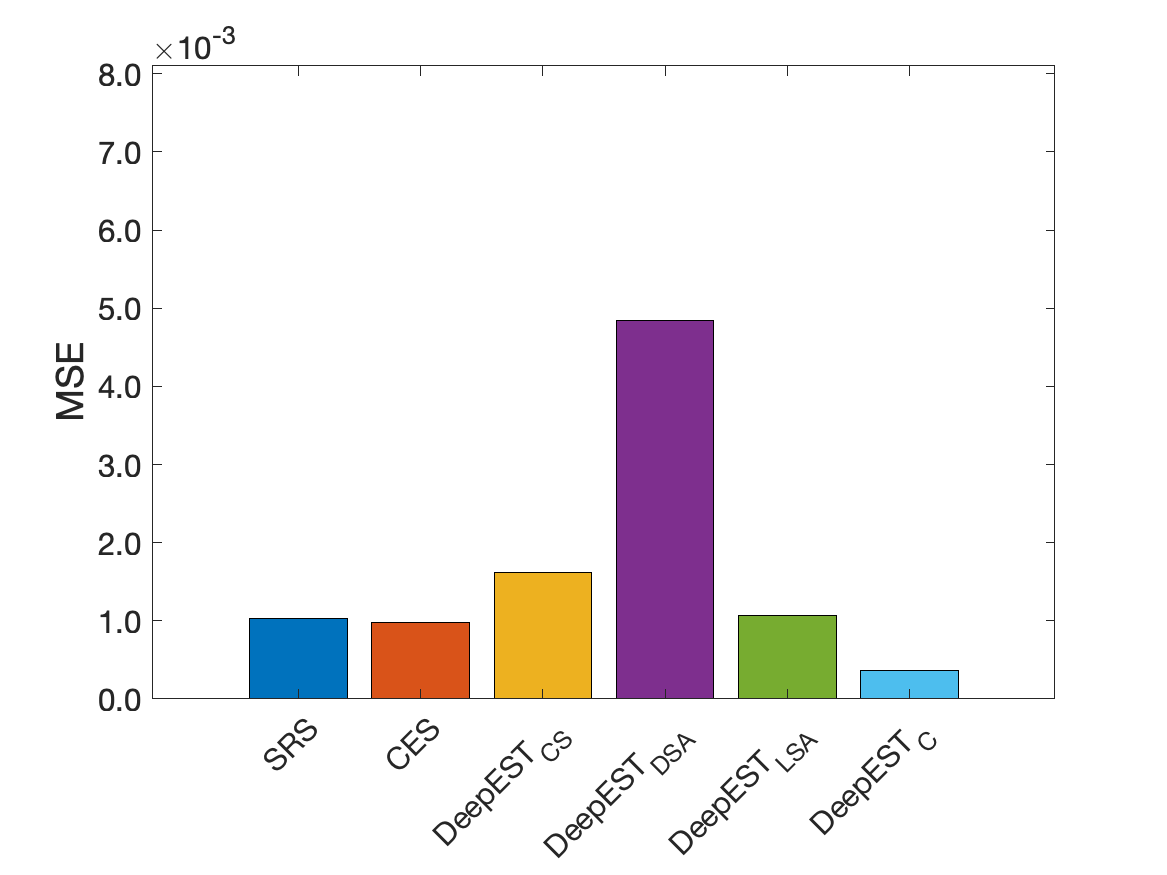}
		\label{3MSE}\hspace{-0,5cm}}
	\subfloat[Subject S4 (VGG16, CIFAR10)]{\includegraphics[width=0.3\textwidth]{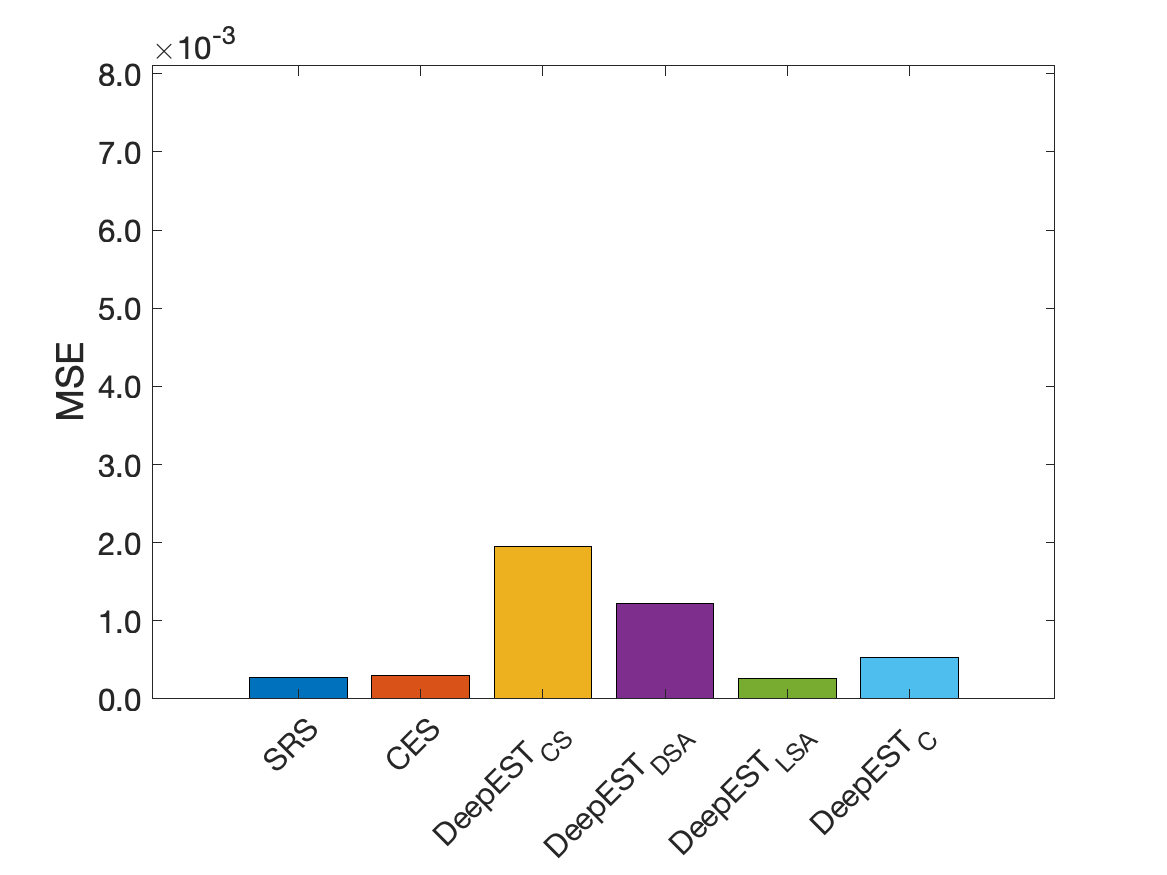}
		\label{4MSE}\hspace{-0,5cm}}
	\subfloat[Subject S5 (VGG16, CIFAR100)]{\includegraphics[width=0.3\textwidth]{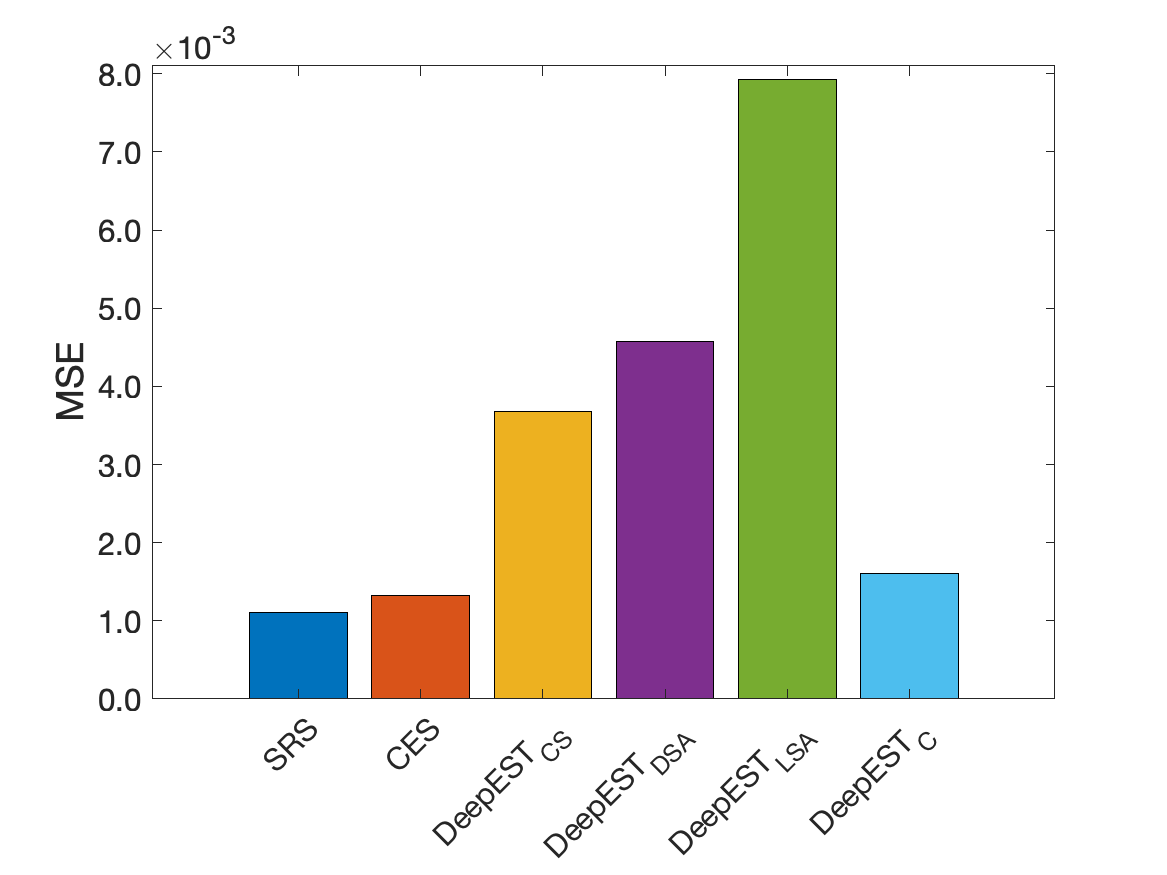}
		\label{5MSE}}
	\caption{RQ1 (effectiveness): Mean Squared Error of estimates}
\vspace{-4pt}
	\label{fig:MSE}
\end{figure*}

\begin{table*}[h!]
	\centering
	\caption{RQ1 (effectiveness): Mean and standard deviation ($\sigma$) of the number of failing examples detected}
	\label{tab:numfp}
	\def\arraystretch{1.01}
	\setlength{\tabcolsep}{5.5pt}
	\begin{tabular}{l|r|c|r|c|r|c|r|c|r|c|r|c|r}
		\toprule
		\multirow{3}{*}{\textbf{Subject}} & \multicolumn{2}{c|}{\textbf{SRS}}& \multicolumn{2}{c|}{\textbf{CES}} & \multicolumn{2}{c|}{\textcolor{blue}{\textbf{\approach{}$_{CS}$}}} & \multicolumn{2}{c|}{\textcolor{blue}{\textbf{\approach{}$_{DSA}$}}} & \multicolumn{2}{c|}{\textcolor{blue}{\textbf{\approach{}$_{LSA}$}}}  & \multicolumn{2}{c|}{\textcolor{blue}{\textbf{\approach{}$_{C}$}}} & \multirow{2}{*}{\begin{tabular}[r]{@{}r@{}}\textbf{Total} \\ \textbf{failing} \\ \textbf{examples}\end{tabular}} \\
		& \begin{tabular}[c]{@{}c@{}}\textbf{Mean \#} \\ \%\end{tabular} & $\sigma$ & \begin{tabular}[c]{@{}c@{}}\textbf{Mean \#} \\ \%\end{tabular} & $\sigma$ & \begin{tabular}[c]{@{}c@{}}\textbf{Mean \#} \\ \%\end{tabular} & $\sigma$ & \begin{tabular}[c]{@{}c@{}}\textbf{Mean \#} \\ \%\end{tabular} & $\sigma$ & \begin{tabular}[c]{@{}c@{}}\textbf{Mean \#} \\ \%\end{tabular} & $\sigma$ & \begin{tabular}[c]{@{}c@{}}\textbf{Mean \#} \\ \%\end{tabular} & $\sigma$ &  \\ \hline
		\begin{tabular}[l]{@{}l@{}}\textbf{S1} \\ (CN5, MNIST) \end{tabular} & %
		\begin{tabular}[c]{@{}c@{}}1.90 \\ \begin{tikzpicture}
			\begin{axis}[xbar stacked, 
			ticks=none, 
			width=0.12\textwidth, 
			bar width=2mm, xmin=0, xmax=100,  y=2mm, enlarge y limits={abs=0.525}]
			\addplot coordinates
			{(2.00,0)}; 
			\end{axis}  
			\pgfresetboundingbox
			\path
			(current axis.south west) -- ++(-0.001in,-0.001in)
			rectangle (current axis.north east) -- ++(0.03in,0.03in);
			\end{tikzpicture} \end{tabular} & 1.52  
		&%
		\begin{tabular}[c]{@{}c@{}}2.23 \\ \begin{tikzpicture}
			\begin{axis}[xbar stacked, 
			ticks=none, 
			width=0.12\textwidth, 
			bar width=2mm, xmin=0, xmax=100,  y=2mm, enlarge y limits={abs=0.525}]
			\addplot coordinates
			{(2.35,0)}; 
			\end{axis}  
			\pgfresetboundingbox
			\path
			(current axis.south west) -- ++(-0.001in,-0.001in)
			rectangle (current axis.north east) -- ++(0.03in,0.03in);
			\end{tikzpicture} \end{tabular} & 1.57  
		& \begin{tabular}[c]{@{}c@{}}44.57 \\ \begin{tikzpicture}
			\begin{axis}[xbar stacked, 
			ticks=none, 
			width=0.12\textwidth, 
			bar width=2mm, xmin=0, xmax=100,  y=2mm, enlarge y limits={abs=0.525}]
			\addplot coordinates
			{(46.91,0)}; 
			\end{axis}  
			\pgfresetboundingbox
			\path
			(current axis.south west) -- ++(-0.001in,-0.001in)
			rectangle (current axis.north east) -- ++(0.03in,0.03in);
			\end{tikzpicture} \end{tabular} & 0.68  
		& \begin{tabular}[c]{@{}c@{}}30.17 \\ \begin{tikzpicture}
			\begin{axis}[xbar stacked, 
			ticks=none, 
			width=0.12\textwidth, 
			bar width=2mm, xmin=0, xmax=100,  y=2mm, enlarge y limits={abs=0.525}]
			\addplot coordinates
			{(31.75,0)}; 
			\end{axis}  
			\pgfresetboundingbox
			\path
			(current axis.south west) -- ++(-0.001in,-0.001in)
			rectangle (current axis.north east) -- ++(0.03in,0.03in);
			\end{tikzpicture} \end{tabular} & 4.03  
		& \begin{tabular}[c]{@{}c@{}}22.43 \\ \begin{tikzpicture}
			\begin{axis}[xbar stacked, 
			ticks=none, 
			width=0.12\textwidth, 
			bar width=2mm, xmin=0, xmax=100,  y=2mm, enlarge y limits={abs=0.525}]
			\addplot coordinates
			{(23.61,0)}; 
			\end{axis}  
			\pgfresetboundingbox
			\path
			(current axis.south west) -- ++(-0.001in,-0.001in)
			rectangle (current axis.north east) -- ++(0.03in,0.03in);
			\end{tikzpicture} \end{tabular} & 3.29 
		&  \begin{tabular}[c]{@{}c@{}}11.93 \\ \begin{tikzpicture}
			\begin{axis}[xbar stacked, 
			ticks=none, 
			width=0.12\textwidth, 
			bar width=2mm, xmin=0, xmax=100,  y=2mm, enlarge y limits={abs=0.525}]
			\addplot coordinates
			{(12.56,0)}; 
			\end{axis}  
			\pgfresetboundingbox
			\path
			(current axis.south west) -- ++(-0.001in,-0.001in)
			rectangle (current axis.north east) -- ++(0.03in,0.03in);
			\end{tikzpicture} \end{tabular} & 3.06 
		& 95 \\ \cline{1-14}
\begin{tabular}[l]{@{}l@{}}\textbf{S2} \\ (LeNet5, MNIST) \end{tabular} & %
		\begin{tabular}[c]{@{}c@{}}2.77 \\ \begin{tikzpicture}
			\begin{axis}[xbar stacked, 
			ticks=none, 
			width=0.12\textwidth, 
			bar width=2mm, xmin=0, xmax=100,  y=2mm, enlarge y limits={abs=0.525}]
			\addplot coordinates
			{(2.10,0)}; 
			\end{axis}  
			\pgfresetboundingbox
			\path
			(current axis.south west) -- ++(-0.001in,-0.001in)
			rectangle (current axis.north east) -- ++(0.03in,0.03in);
			\end{tikzpicture} \end{tabular} & 2.06  
		&%
		\begin{tabular}[c]{@{}c@{}}2.03 \\ \begin{tikzpicture}
			\begin{axis}[xbar stacked, 
			ticks=none, 
			width=0.12\textwidth, 
			bar width=2mm, xmin=0, xmax=100,  y=2mm, enlarge y limits={abs=0.525}]
			\addplot coordinates
			{(1.54,0)}; 
			\end{axis}  
			\pgfresetboundingbox
			\path
			(current axis.south west) -- ++(-0.001in,-0.001in)
			rectangle (current axis.north east) -- ++(0.03in,0.03in);
			\end{tikzpicture} \end{tabular} & 1.16  
		& \begin{tabular}[c]{@{}c@{}}65.13 \\ \begin{tikzpicture}
			\begin{axis}[xbar stacked, 
			ticks=none, 
			width=0.12\textwidth, 
			bar width=2mm, xmin=0, xmax=100,  y=2mm, enlarge y limits={abs=0.525}]
			\addplot coordinates
			{(49.34,0)}; 
			\end{axis}  
			\pgfresetboundingbox
			\path
			(current axis.south west) -- ++(-0.001in,-0.001in)
			rectangle (current axis.north east) -- ++(0.03in,0.03in);
			\end{tikzpicture} \end{tabular} & 0.86 
		& \begin{tabular}[c]{@{}c@{}}37.57 \\ \begin{tikzpicture}
			\begin{axis}[xbar stacked, 
			ticks=none, 
			width=0.12\textwidth, 
			bar width=2mm, xmin=0, xmax=100,  y=2mm, enlarge y limits={abs=0.525}]
			\addplot coordinates
			{(28.46,0)}; 
			\end{axis}  
			\pgfresetboundingbox
			\path
			(current axis.south west) -- ++(-0.001in,-0.001in)
			rectangle (current axis.north east) -- ++(0.03in,0.03in);
			\end{tikzpicture} \end{tabular} & 5.12 
		& \begin{tabular}[c]{@{}c@{}}24.10 \\ \begin{tikzpicture}
			\begin{axis}[xbar stacked, 
			ticks=none, 
			width=0.12\textwidth, 
			bar width=2mm, xmin=0, xmax=100,  y=2mm, enlarge y limits={abs=0.525}]
			\addplot coordinates
			{(18.26,0)}; 
			\end{axis}  
			\pgfresetboundingbox
			\path
			(current axis.south west) -- ++(-0.001in,-0.001in)
			rectangle (current axis.north east) -- ++(0.03in,0.03in);
			\end{tikzpicture} \end{tabular} & 3.74 
		&  \begin{tabular}[c]{@{}c@{}}24.07 \\ \begin{tikzpicture}
			\begin{axis}[xbar stacked, 
			ticks=none, 
			width=0.12\textwidth, 
			bar width=2mm, xmin=0, xmax=100,  y=2mm, enlarge y limits={abs=0.525}]
			\addplot coordinates
			{(18.23,0)}; 
			\end{axis}  
			\pgfresetboundingbox
			\path
			(current axis.south west) -- ++(-0.001in,-0.001in)
			rectangle (current axis.north east) -- ++(0.03in,0.03in);
			\end{tikzpicture} \end{tabular} & 4.15 
		& 132 \\ \hline
\begin{tabular}[l]{@{}l@{}}\textbf{S3} \\ (CN12, CIFAR10) \end{tabular} & %
		\begin{tabular}[c]{@{}c@{}}37.93 \\ \begin{tikzpicture}
			\begin{axis}[xbar stacked, 
			ticks=none, 
			width=0.12\textwidth, 
			bar width=2mm, xmin=0, xmax=100,  y=2mm, enlarge y limits={abs=0.525}]
			\addplot coordinates
			{(1.96,0)}; 
			\end{axis}  
			\pgfresetboundingbox
			\path
			(current axis.south west) -- ++(-0.001in,-0.001in)
			rectangle (current axis.north east) -- ++(0.03in,0.03in);
			\end{tikzpicture} \end{tabular} & 6.47 
		&%
		\begin{tabular}[c]{@{}c@{}}33.77 \\ \begin{tikzpicture}
			\begin{axis}[xbar stacked, 
			ticks=none, 
			width=0.12\textwidth, 
			bar width=2mm, xmin=0, xmax=100,  y=2mm, enlarge y limits={abs=0.525}]
			\addplot coordinates
			{(1.75,0)}; 
			\end{axis}  
			\pgfresetboundingbox
			\path
			(current axis.south west) -- ++(-0.001in,-0.001in)
			rectangle (current axis.north east) -- ++(0.03in,0.03in);
			\end{tikzpicture} \end{tabular} & 3.92 
		& \begin{tabular}[c]{@{}c@{}}106.10 \\ \begin{tikzpicture}
			\begin{axis}[xbar stacked, 
			ticks=none, 
			width=0.12\textwidth, 
			bar width=2mm, xmin=0, xmax=100,  y=2mm, enlarge y limits={abs=0.525}]
			\addplot coordinates
			{(5.49,0)}; 
			\end{axis}  
			\pgfresetboundingbox
			\path
			(current axis.south west) -- ++(-0.001in,-0.001in)
			rectangle (current axis.north east) -- ++(0.03in,0.03in);
			\end{tikzpicture} \end{tabular} & 7.26 
		& \begin{tabular}[c]{@{}c@{}}98.33 \\ \begin{tikzpicture}
			\begin{axis}[xbar stacked, 
			ticks=none, 
			width=0.12\textwidth, 
			bar width=2mm, xmin=0, xmax=100,  y=2mm, enlarge y limits={abs=0.525}]
			\addplot coordinates
			{(5.08,0)}; 
			\end{axis}  
			\pgfresetboundingbox
			\path
			(current axis.south west) -- ++(-0.001in,-0.001in)
			rectangle (current axis.north east) -- ++(0.03in,0.03in);
			\end{tikzpicture} \end{tabular} & 4.62 
		& \begin{tabular}[c]{@{}c@{}}38.07 \\ \begin{tikzpicture}
			\begin{axis}[xbar stacked, 
			ticks=none, 
			width=0.12\textwidth, 
			bar width=2mm, xmin=0, xmax=100,  y=2mm, enlarge y limits={abs=0.525}]
			\addplot coordinates
			{(1.97,0)}; 
			\end{axis}  
			\pgfresetboundingbox
			\path
			(current axis.south west) -- ++(-0.001in,-0.001in)
			rectangle (current axis.north east) -- ++(0.03in,0.03in);
			\end{tikzpicture} \end{tabular} & 6.61 
		&  \begin{tabular}[c]{@{}c@{}}70.77 \\ \begin{tikzpicture}
			\begin{axis}[xbar stacked, 
			ticks=none, 
			width=0.12\textwidth, 
			bar width=2mm, xmin=0, xmax=100,  y=2mm, enlarge y limits={abs=0.525}]
			\addplot coordinates
			{(3.66,0)}; 
			\end{axis}  
			\pgfresetboundingbox
			\path
			(current axis.south west) -- ++(-0.001in,-0.001in)
			rectangle (current axis.north east) -- ++(0.03in,0.03in);
			\end{tikzpicture} \end{tabular} & 7.09 
		& 1,934 \\ \cline{1-14}
\begin{tabular}[l]{@{}l@{}}\textbf{S4} \\ (VGG16, CIFAR10) \end{tabular} & %
		\begin{tabular}[c]{@{}c@{}}12.60 \\ \begin{tikzpicture}
			\begin{axis}[xbar stacked, 
			ticks=none, 
			width=0.12\textwidth, 
			bar width=2mm, xmin=0, xmax=100,  y=2mm, enlarge y limits={abs=0.525}]
			\addplot coordinates
			{(1.97,0)}; 
			\end{axis}  
			\pgfresetboundingbox
			\path
			(current axis.south west) -- ++(-0.001in,-0.001in)
			rectangle (current axis.north east) -- ++(0.03in,0.03in);
			\end{tikzpicture} \end{tabular} & 3.33 
		&%
		\begin{tabular}[c]{@{}c@{}}13.03 \\ \begin{tikzpicture}
			\begin{axis}[xbar stacked, 
			ticks=none, 
			width=0.12\textwidth, 
			bar width=2mm, xmin=0, xmax=100,  y=2mm, enlarge y limits={abs=0.525}]
			\addplot coordinates
			{(2.03,0)}; 
			\end{axis}  
			\pgfresetboundingbox
			\path
			(current axis.south west) -- ++(-0.001in,-0.001in)
			rectangle (current axis.north east) -- ++(0.03in,0.03in);
			\end{tikzpicture} \end{tabular} & 3.49 
		& \begin{tabular}[c]{@{}c@{}}85.00 \\ \begin{tikzpicture}
			\begin{axis}[xbar stacked, 
			ticks=none, 
			width=0.12\textwidth, 
			bar width=2mm, xmin=0, xmax=100,  y=2mm, enlarge y limits={abs=0.525}]
			\addplot coordinates
			{(13.26,0)}; 
			\end{axis}  
			\pgfresetboundingbox
			\path
			(current axis.south west) -- ++(-0.001in,-0.001in)
			rectangle (current axis.north east) -- ++(0.03in,0.03in);
			\end{tikzpicture} \end{tabular} & 4.86 
		& \begin{tabular}[c]{@{}c@{}}16.80 \\ \begin{tikzpicture}
			\begin{axis}[xbar stacked, 
			ticks=none, 
			width=0.12\textwidth, 
			bar width=2mm, xmin=0, xmax=100,  y=2mm, enlarge y limits={abs=0.525}]
			\addplot coordinates
			{(2.62,0)}; 
			\end{axis}  
			\pgfresetboundingbox
			\path
			(current axis.south west) -- ++(-0.001in,-0.001in)
			rectangle (current axis.north east) -- ++(0.03in,0.03in);
			\end{tikzpicture} \end{tabular} & 3.02 
		& \begin{tabular}[c]{@{}c@{}}14.77 \\ \begin{tikzpicture}
			\begin{axis}[xbar stacked, 
			ticks=none, 
			width=0.12\textwidth, 
			bar width=2mm, xmin=0, xmax=100,  y=2mm, enlarge y limits={abs=0.525}]
			\addplot coordinates
			{(2.30,0)}; 
			\end{axis}  
			\pgfresetboundingbox
			\path
			(current axis.south west) -- ++(-0.001in,-0.001in)
			rectangle (current axis.north east) -- ++(0.03in,0.03in);
			\end{tikzpicture} \end{tabular} & 3.14 
		&  \begin{tabular}[c]{@{}c@{}}24.90 \\ \begin{tikzpicture}
			\begin{axis}[xbar stacked, 
			ticks=none, 
			width=0.12\textwidth, 
			bar width=2mm, xmin=0, xmax=100,  y=2mm, enlarge y limits={abs=0.525}]
			\addplot coordinates
			{(3.88,0)}; 
			\end{axis}  
			\pgfresetboundingbox
			\path
			(current axis.south west) -- ++(-0.001in,-0.001in)
			rectangle (current axis.north east) -- ++(0.03in,0.03in);
			\end{tikzpicture} \end{tabular} & 4.97 
		& 641 \\ \hline
\begin{tabular}[l]{@{}l@{}}\textbf{S5} \\ (VGG16, CIFAR100) \end{tabular} & %
		\begin{tabular}[c]{@{}c@{}}57.33 \\ \begin{tikzpicture}
			\begin{axis}[xbar stacked, 
			ticks=none, 
			width=0.12\textwidth, 
			bar width=2mm, xmin=0, xmax=100,  y=2mm, enlarge y limits={abs=0.525}]
			\addplot coordinates
			{(1.94,0)}; 
			\end{axis}  
			\pgfresetboundingbox
			\path
			(current axis.south west) -- ++(-0.001in,-0.001in)
			rectangle (current axis.north east) -- ++(0.03in,0.03in);
			\end{tikzpicture} \end{tabular} & 6.53 
		&%
		\begin{tabular}[c]{@{}c@{}}55.77 \\ \begin{tikzpicture}
			\begin{axis}[xbar stacked, 
			ticks=none, 
			width=0.12\textwidth, 
			bar width=2mm, xmin=0, xmax=100,  y=2mm, enlarge y limits={abs=0.525}]
			\addplot coordinates
			{(1.89,0)}; 
			\end{axis}  
			\pgfresetboundingbox
			\path
			(current axis.south west) -- ++(-0.001in,-0.001in)
			rectangle (current axis.north east) -- ++(0.03in,0.03in);
			\end{tikzpicture} \end{tabular} & 6.62 
		& \begin{tabular}[c]{@{}c@{}}131.17 \\ \begin{tikzpicture}
			\begin{axis}[xbar stacked, 
			ticks=none, 
			width=0.12\textwidth, 
			bar width=2mm, xmin=0, xmax=100,  y=2mm, enlarge y limits={abs=0.525}]
			\addplot coordinates
			{(4.44,0)}; 
			\end{axis}  
			\pgfresetboundingbox
			\path
			(current axis.south west) -- ++(-0.001in,-0.001in)
			rectangle (current axis.north east) -- ++(0.03in,0.03in);
			\end{tikzpicture} \end{tabular} & 7.35 
		& \begin{tabular}[c]{@{}c@{}}78.20 \\ \begin{tikzpicture}
			\begin{axis}[xbar stacked, 
			ticks=none, 
			width=0.12\textwidth, 
			bar width=2mm, xmin=0, xmax=100,  y=2mm, enlarge y limits={abs=0.525}]
			\addplot coordinates
			{(2.65,0)}; 
			\end{axis}  
			\pgfresetboundingbox
			\path
			(current axis.south west) -- ++(-0.001in,-0.001in)
			rectangle (current axis.north east) -- ++(0.03in,0.03in);
			\end{tikzpicture} \end{tabular} & 6.62 
		& \begin{tabular}[c]{@{}c@{}}67.47 \\ \begin{tikzpicture}
			\begin{axis}[xbar stacked, 
			ticks=none, 
			width=0.12\textwidth, 
			bar width=2mm, xmin=0, xmax=100,  y=2mm, enlarge y limits={abs=0.525}]
			\addplot coordinates
			{(2.29,0)}; 
			\end{axis}  
			\pgfresetboundingbox
			\path
			(current axis.south west) -- ++(-0.001in,-0.001in)
			rectangle (current axis.north east) -- ++(0.03in,0.03in);
			\end{tikzpicture} \end{tabular} & 3.68 
		&  \begin{tabular}[c]{@{}c@{}}108.70 \\ \begin{tikzpicture}
			\begin{axis}[xbar stacked, 
			ticks=none, 
			width=0.12\textwidth, 
			bar width=2mm, xmin=0, xmax=100,  y=2mm, enlarge y limits={abs=0.525}]
			\addplot coordinates
			{(3.68,0)}; 
			\end{axis}  
			\pgfresetboundingbox
			\path
			(current axis.south west) -- ++(-0.001in,-0.001in)
			rectangle (current axis.north east) -- ++(0.03in,0.03in);
			\end{tikzpicture} \end{tabular} & 4.45 
		& 2,952 \\
		\bottomrule
	\end{tabular}
\vspace{-4pt}
\end{table*}

\subsection{RQ1: effectiveness}
Figure \ref{fig:MSE} plots the MSE of the estimated accuracy.   
The techniques exhibit comparable performances, with \approach{}$_C$ being the best one for 3 of the 5 subjects. 
CES has good performance in terms of MSE, it is the second technique in 3 cases. 
Considering the single variables: confidence, DSA or LSA %
lead to results slightly more variable over the subjects 
-- an aspect explored in RQ3. 
The SRS case is interesting, too: it is never the worst approach and is the best in one case. %

\begin{table*}[htp]
	\caption{Pairwise comparison of techniques. A value of $\rho_{R,C}$$>$$1$ means the technique on the row has a greater precision than that on the column. Similarly for the relative number of detected failures $\pi_{R,C}$}
		\label{tabellone}
		\centering
	\subfloat[Subject S1 (CN5, MNIST)]{
		\centering
		\def\arraystretch{1.1}
		\setlength{\tabcolsep}{4.5pt}
		\begin{tabular}{c|c|c|c|c|c|c|r}
			\toprule
			& \multicolumn{2}{c|}{} & \multicolumn{4}{c|}{\textcolor{blue}{DeepEST}} & \\\cline{4-7}
			& \multicolumn{2}{c|}{\textit{row vs col}} & \textcolor{cyan}{$CS$} & \textcolor{blue}{$DSA$} & \textcolor{orange}{$LSA$} & \textcolor{red}{$C$} & {SRS} \\ \hline
			& \multirow{2}{*}{{CES}} & \textit{$\rho_{R,C}$} & 0.0501 & \textcolor{blue}{0.0740} & \textcolor{orange}{0.0996} & \textcolor{red}{0.1872} & 1.1754 \\ \cdashline{3-8} 
			& & \textit{$\pi_{R,C}$} & 1.0987 & \textcolor{blue}{0.6790} & \textcolor{orange}{0.3564} & \textcolor{red}{0.1136} & 0.8929 \\ \hline 
			\multirow{8}{*}{\STAB{\rotatebox[origin=c]{90}{\textcolor{blue}{DeepEST}}}}
			& \multirow{2}{*}{\textcolor{cyan}{\textit{$CS$}}} & \textit{$\rho_{R,C}$} & \multirow{2}{*}{-} & 1.4773 & 1.9866 & 3.7346 & 23.4561 \\ \cdashline{3-3} \cdashline{5-8} 
			& & \textit{$\pi_{R,C}$} & & 0.6180 & 0.3244 & 0.1034 & 0.8127 \\ \cdashline{3-3} \cline{2-8} 
			& \multirow{2}{*}{\textcolor{blue}{$DSA$}} & \textit{$\rho_{R,C}$} & \multirow{2}{*}{-} & \multirow{2}{*}{-} & 1.3447 & 2.5279 & \textcolor{blue}{15.8772} \\ \cdashline{3-3} \cdashline{6-8} 
			& & \textit{$\pi_{R,C}$} & & & 0.5249 & 0.1673 & \textcolor{blue}{1.3150} \\ \cdashline{2-2} \cline{2-8} 
			& \multirow{2}{*}{ \textcolor{orange}{$LSA$} } & \textit{$\rho_{R,C}$} & \multirow{2}{*}{-} & \multirow{2}{*}{-} & \multirow{2}{*}{-} & 1.8799 & \textcolor{orange}{11.8070} \\ \cdashline{3-3} \cdashline{7-8} 
			& & \textit{$\pi_{R,C}$} & & & & 0.3187 & \textcolor{orange}{2.5054} \\ \cdashline{3-3} \cline{2-8}
			& \multirow{2}{*}{ \textcolor{red}{$C$}} & \textit{$\rho_{R,C}$} & \multirow{2}{*}{-} & \multirow{2}{*}{-} & \multirow{2}{*}{-} & \multirow{2}{*}{-} & \textcolor{red}{6.2807} \\ \cdashline{3-3} \cdashline{8-8} 
			& & \textit{$\pi_{R,C}$} & & & & & \textcolor{red}{7.8624} \\ 
			\bottomrule
		\end{tabular}
		\label{tab:vsMNISTconv5}
	}
\hspace{0.1cm}
	\subfloat[Subject S2 (LeNet5, MNIST)]{
		\centering
		\def\arraystretch{1.1}
		\setlength{\tabcolsep}{5pt}
		\begin{tabular}{c|c|c|c|c|c|c|r}
			\toprule
			& \multicolumn{2}{c|}{} & \multicolumn{4}{c|}{\textcolor{blue}{DeepEST}} & \\\cline{4-7}
			& \multicolumn{2}{c|}{\textit{row vs col}} & \textcolor{cyan}{$CS$} & \textcolor{blue}{$DSA$} & \textcolor{orange}{$LSA$} & \textcolor{red}{$C$} & {SRS} \\ \hline
			& \multirow{2}{*}{{CES}} & \textit{$\rho_{R,C}$} & 0.0312 & 0.0541 & 0.0844 & \textcolor{red}{0.0845} & 0.7349 \\ \cdashline{3-8} 
			& & \textit{$\pi_{R,C}$} & 3.9040 & 2.9238 & 1.4523 & \textcolor{red}{0.1473} & 2.4766 \\ \hline 
			\multirow{8}{*}{\STAB{\rotatebox[origin=c]{90}{\textcolor{blue}{DeepEST}}}}
			& \multirow{2}{*}{\textcolor{cyan}{\textit{$CS$}}} & \textit{$\rho_{R,C}$} & \multirow{2}{*}{-} & 1.7338 & 2.7026 & 2.7064 & 23.5422 \\ \cdashline{3-3} \cdashline{5-8} 
			& & \textit{$\pi_{R,C}$} &  & 0.7489 & 0.3720 & 0.0377 & 0.6344 \\ \cdashline{3-3} \cline{2-8}  
			& \multirow{2}{*}{\textcolor{blue}{$DSA$}} & \textit{$\rho_{R,C}$} & \multirow{2}{*}{-} & \multirow{2}{*}{-} & 1.5588 & 1.5609 & 13.5783 \\ \cdashline{3-3} \cdashline{6-8} 
			& & \textit{$\pi_{R,C}$} &  &  & 0.4967 & 0.0504 & 0.8470 \\ \cdashline{3-3} \cline{2-8} 
			& \multirow{2}{*}{\textcolor{orange}{$LSA$} } & \textit{$\rho_{R,C}$} & \multirow{2}{*}{-} & \multirow{2}{*}{-} & \multirow{2}{*}{-} & 1.0014 & \textcolor{orange}{8.7108} \\ \cdashline{3-3} \cdashline{7-8} 
			& & \textit{$\pi_{R,C}$} &  &  &  & 0.1014 & \textcolor{orange}{1.7053} \\ \cdashline{3-3} \cline{2-8} 
			& \multirow{2}{*}{\textcolor{red}{$C$}} & \textit{$\rho_{R,C}$} & \multirow{2}{*}{-} & \multirow{2}{*}{-} & \multirow{2}{*}{-} & \multirow{2}{*}{-} & \textcolor{red}{8.6988} \\ \cdashline{3-3} \cdashline{8-8} 
			& & \textit{$\pi_{R,C}$} &  &  &  &  & \textcolor{red}{16.8140} \\ 
			\bottomrule
		\end{tabular}
		\label{tab:vsMNISTlen5}
	}\\
\vspace{-2pt}
	\subfloat[Subject S3 (CN12, CIFAR10)]{
		\centering
		\def\arraystretch{1.1}
		\setlength{\tabcolsep}{5pt}
		\begin{tabular}{c|c|c|c|c|c|c|r}
			\toprule
			& \multicolumn{2}{c|}{} & \multicolumn{4}{c|}{\textcolor{blue}{DeepEST}} & \\\cline{4-7}
			& \multicolumn{2}{c|}{\textit{row vs col}} & \textcolor{cyan}{$CS$} & \textcolor{blue}{$DSA$} & \textcolor{orange}{$LSA$} & \textcolor{red}{$C$} & {SRS} \\ \hline
			& \multirow{2}{*}{{CES}} & \textit{$\rho_{R,C}$} & 0.3183 & 0.3434 & 0.8870 & \textcolor{red}{0.4772} & {0.8902} \\ \cdashline{3-8} 
			& & \textit{$\pi_{R,C}$} & 1.6669 & 4.9728 & 1.0967 & \textcolor{red}{0.3741} & 1.0522 \\ \hline 
			\multirow{8}{*}{\STAB{\rotatebox[origin=c]{90}{\textcolor{blue}{DeepEST}}}}
			& \multirow{2}{*}{\textcolor{cyan}{\textit{$CS$}}} & \textit{$\rho_{R,C}$} & \multirow{2}{*}{-} & \textcolor{cyan}{1.0790} & 2.7872 & 1.4993 & 2.7970 \\ \cdashline{3-3} \cdashline{5-8} 
			& & \textit{$\pi_{R,C}$} &  & \textcolor{cyan}{2.9833} & 0.6579 & 0.2244 & 0.6312 \\ \cdashline{3-3} \cline{2-8}  
			& \multirow{2}{*}{\textcolor{blue}{$DSA$}} & \textit{$\rho_{R,C}$} & \multirow{2}{*}{-} & \multirow{2}{*}{-} & 2.5832 & 1.3895 & 2.5923 \\ \cdashline{3-3} \cdashline{6-8} 
			& & \textit{$\pi_{R,C}$} &  &  & 0.2205 & 0.0752 & 0.2116 \\ \cdashline{2-2} \cline{2-8} 
			& \multirow{2}{*}{\textcolor{orange}{$LSA$} } & \textit{$\rho_{R,C}$} & \multirow{2}{*}{-} & \multirow{2}{*}{-} & \multirow{2}{*}{-} & \textcolor{red}{0.5379} & 1.0035 \\ \cdashline{3-3} \cdashline{7-8} 
			& & \textit{$\pi_{R,C}$} &  &  &  & \textcolor{red}{0.3411} & 0.9594 \\ \cdashline{3-3} \cline{2-8}   
			& \multirow{2}{*}{\textcolor{red}{$C$}} & \textit{$\rho_{R,C}$} & \multirow{2}{*}{-} & \multirow{2}{*}{-} & \multirow{2}{*}{-} & \multirow{2}{*}{-} & \textcolor{red}{1.8656} \\ \cdashline{3-3} \cdashline{8-8} 
			& & \textit{$\pi_{R,C}$} &  &  &  &  & \textcolor{red}{2.8126} \\ 
			\bottomrule
		\end{tabular}
		\label{tab:vsCIFAR10conv12}
	}
\hspace{0.15cm}
	\subfloat[Subject S4 (VGG16, CIFAR10)]{
		\centering
		\def\arraystretch{1.1}
		\setlength{\tabcolsep}{5pt}
		\begin{tabular}{c|c|c|c|c|c|c|r}
			\toprule
			& \multicolumn{2}{c|}{} & \multicolumn{4}{c|}{\textcolor{blue}{DeepEST}} & \\\cline{4-7}
			& \multicolumn{2}{c|}{\textit{row vs col}} & \textcolor{cyan}{$CS$} & \textcolor{blue}{$DSA$} & \textcolor{orange}{$LSA$} & \textcolor{red}{$C$} & {SRS} \\ \hline
			& \multirow{2}{*}{{CES}} & \textit{$\rho_{R,C}$} & 0.1533 & 0.7758 & \textcolor{orange}{0.8826} & 0.5234 & {1.0344} \\ \cdashline{3-8} 
			& & \textit{$\pi_{R,C}$} & 6.5937 & 4.1503 & \textcolor{orange}{0.8772} & 1.7910 & 0.9106 \\ \hline 
			\multirow{8}{*}{\STAB{\rotatebox[origin=c]{90}{\textcolor{blue}{DeepEST}}}}
			& \multirow{2}{*}{\textcolor{cyan}{\textit{$CS$}}} & \textit{$\rho_{R,C}$} & \multirow{2}{*}{-} & 5.0595 & 5.7562 & 3.4137 & 6.7460 \\ \cdashline{3-3} \cdashline{5-8} 
			& & \textit{$\pi_{R,C}$} &  & 0.6294 & 0.1330 & 0.2716 & 0.1381 \\ \cdashline{2-2} \cline{2-8} 
			& \multirow{2}{*}{\textcolor{blue}{$DSA$}} & \textit{$\rho_{R,C}$} & \multirow{2}{*}{-} & \multirow{2}{*}{-} & 1.1377 & \textcolor{red}{0.6747} & 1.3333 \\ \cdashline{3-3} \cdashline{6-8} 
			& & \textit{$\pi_{R,C}$} &  &  & 0.2114 & \textcolor{red}{0.4315} & 0.2194 \\ \cdashline{2-2} \cline{2-8}
			& \multirow{2}{*}{\textcolor{orange}{$LSA$} } & \textit{$\rho_{R,C}$} & \multirow{2}{*}{-} & \multirow{2}{*}{-} & \multirow{2}{*}{-} & 0.5930 & \textcolor{orange}{1.1720} \\ \cdashline{3-3} \cdashline{7-8} 
			& & \textit{$\pi_{R,C}$} &  &  &  & 2.0417 & \textcolor{orange}{1.0380} \\ \cdashline{3-3} \cline{2-8}  
			& \multirow{2}{*}{\textcolor{red}{$C$}} & \textit{$\rho_{R,C}$} & \multirow{2}{*}{-} & \multirow{2}{*}{-} & \multirow{2}{*}{-} & \multirow{2}{*}{-} & 1.9762 \\ \cdashline{3-3} \cdashline{8-8} 
			& & \textit{$\pi_{R,C}$} &  &  &  &  & 0.5084 \\ 
			\bottomrule
		\end{tabular}
		\label{tab:vsCIFAR10vgg}
	}\\
\vspace{-2pt}
	\subfloat[Subject S5 (VGG16, CIFAR100)]{
		\centering
		\def\arraystretch{1.1}
		\setlength{\tabcolsep}{5pt}
		\begin{tabular}{c|c|c|c|c|c|c|r}
			\toprule
			& \multicolumn{2}{c|}{} & \multicolumn{4}{c|}{\textcolor{blue}{DeepEST}} & \\\cline{4-7}
			& \multicolumn{2}{c|}{\textit{row vs col}} & \textcolor{cyan}{$CS$} & \textcolor{blue}{$DSA$} & \textcolor{orange}{$LSA$} & \textcolor{red}{$C$} & {SRS} \\ \hline
			& \multirow{2}{*}{{CES}} & \textit{$\rho_{R,C}$} & 0.4252 & 0.7131 & 0.8266 & 0.5130 & 0.9727 \\ \cdashline{3-8} 
			& & \textit{$\pi_{R,C}$} & 2.7761 & 3.4999 & 3.5473 & 1.2076 & 0.8323 \\ \hline
			\multirow{8}{*}{\STAB{\rotatebox[origin=c]{90}{\textcolor{blue}{DeepEST}}}}
			& \multirow{2}{*}{\textcolor{cyan}{\textit{$CS$}}} & \textit{$\rho_{R,C}$} & \multirow{2}{*}{-} & \textcolor{cyan}{1.6773} & \textcolor{cyan}{1.9442} & 1.2067 & 2.2878 \\ \cdashline{3-3} \cdashline{5-8} 
			& & \textit{$\pi_{R,C}$} &  & \textcolor{cyan}{1.2607} & \textcolor{cyan}{1.2778} & 0.4350 & 0.2998 \\ \cdashline{2-2} \cline{2-8} 
			& \multirow{2}{*}{\textcolor{blue}{$DSA$}} & \textit{$\rho_{R,C}$} & \multirow{2}{*}{-} & \multirow{2}{*}{-} &  \textcolor{blue}{1.1591} & \textcolor{red}{0.7194} & 1.3640 \\ \cdashline{3-3} \cdashline{6-8} 
			& & \textit{$\pi_{R,C}$} &  &  & \textcolor{blue}{1.0136} & \textcolor{red}{0.3450} & 0.2378 \\ \cdashline{2-2} \cline{2-8}
			& \multirow{2}{*}{\textcolor{orange}{$LSA$} } & \textit{$\rho_{R,C}$} & \multirow{2}{*}{-} & \multirow{2}{*}{-} & \multirow{2}{*}{-} & \textcolor{red}{0.6207} & 1.1767 \\ \cdashline{3-3} \cdashline{7-8} 
			& & \textit{$\pi_{R,C}$} &  &  &  & \textcolor{red}{0.3404} & 0.2346 \\ \cdashline{3-3} \cline{2-8} 
			& \multirow{2}{*}{\textcolor{red}{$C$}} & \textit{$\rho_{R,C}$} & \multirow{2}{*}{-} & \multirow{2}{*}{-} & \multirow{2}{*}{-} & \multirow{2}{*}{-} & 1.8959 \\ \cdashline{3-3} \cdashline{8-8} 
			& & \textit{$\pi_{R,C}$} &  &  &  &  & 0.6892 \\ 
			\bottomrule
		\end{tabular}
		\label{tab:vsCIFAR100vgg}
	}
\hspace{0.25cm}
	\subfloat[Number of wins and losses]{
		\centering
		\def\arraystretch{1.2}
		\setlength{\tabcolsep}{5pt}
		\begin{tabular}{c|c|c|c|c|c|c|c|r}
			\toprule
			& \textit{wins} & & \multicolumn{4}{c|}{\textcolor{blue}{DeepEST}} & & \textbf{Total}\\\cline{4-7}
			& \textit{(r vs c)} & {CES} & \textcolor{cyan}{\textit{{CS}}} & \textcolor{blue}{\textit{DSA}} & \textcolor{orange}{\textit{LSA}} & \textcolor{red}{$C$} & {SRS} & \textbf{wins} \\ \hline
			&{CES} & - & 0 & 0 & 0 & 0 & 0 & 0/25 \\ \hline
			\multirow{4}{*}{\STAB{\rotatebox[origin=c]{90}{\textcolor{blue}{DeepEST}}}}
			& \textcolor{cyan}{\textit{CS}} & 0 & - & \textcolor{cyan}{2} & \textcolor{cyan}{1} & 0 & 0 &  \textcolor{cyan}{3/25} \\ \cline{2-9}
			& \textcolor{blue}{$DSA$} & \textcolor{blue}{1} & 0 & - & \textcolor{blue}{1} & 0 & \textcolor{blue}{1} &  \textcolor{blue}{3/25} \\ \cline{2-9}
			& \textcolor{orange}{\textit{LSA}} & \textcolor{orange}{2} & 0 & 0 & - & 0 & \textcolor{orange}{3} &  \textcolor{orange}{5/25} \\ \cline{2-9}
			& \textcolor{red}{$C$} & \textcolor{red}{3} & 0 & \textcolor{red}{2} & \textcolor{red}{2} & - & \textcolor{red}{3} &  \textcolor{red}{\textbf{10/25}} \\ \hline
			& {SRS} & 1 & 0 & 0 & 0 & 0 & - & 1/25 \\ 
			\bottomrule
		\toprule
			& \textbf{Total} & {CES} & \textcolor{cyan}{$CS$} &  \textcolor{blue}{$DSA$} & \textcolor{orange}{\textit{LSA}} & \textcolor{red}{$C$}& {SRS} & \\ \cline{3-8} 
			& \textbf{losses}& 7/25 & \textcolor{cyan}{\textbf{0/25}} &  \textcolor{blue}{4/25} & \textcolor{orange}{4/25} & \textcolor{red}{\textbf{0/25}} & 7/25\\
			\bottomrule
		\end{tabular} 
		\label{tab:wins}
	}
\vspace{-15pt}
\end{table*}

Table \ref{tab:numfp} %
reports the average number and the standard deviation of the failing examples detected. All variants of \approach{} identify many more failing examples than SRS and CES, even up to a factor of 30x (\approach{}$_{CS}$ $vs$ CES for subject S2) and reaching in some cases (S1 and S2, \approach{}$_{CS}$) almost 50\% of the total number of failing examples in the datasets (last column). The \approach{} algorithm leverages the adaptive sampling to spot clusters of failing examples with relatively few tests (set to 200 for RQ1). Its performance varies depending on the auxiliary information used, but it is always remarkably better than SRS and CES. 

Among the \approach{} variants, 
confidence (\approach{}$_{CS}$) turns out to be the most effective auxiliary variable in detecting failures, showing the best performance for all datasets and models, followed by DSA. 
CES and SRS select the lowest number of mispredicted examples, and are close to each other.

Considering both the failure detection ability and the esti\-mate accuracy,  \approach{}$_C$ -- that combines confidence and DSA, the two best auxiliary variables for failing examples detection - gives a good trade-off,   %
since it provides stable (across subjects) and close-to-true estimates of the accuracy, with many more detected failing examples than CES and SRS. %

Tables \ref{tabellone}(a)--(e) show the results of the pairwise comparison of the techniques. %
For the four \approach{} variants, rows and columns headings list (in blue) the name of the auxiliary variables. 
The evaluation metrics are the ratio $\rho$ of the failing examples and the relative precision $\pi$ of the estimators. 
Values of $\rho$ or $\pi$ greater (lower) than 1 mean the technique on the row (column) has better performance. 
If a technique is better than the other in a pair for both metrics (values coloured in the table), we say that it \textit{wins}.

\begin{figure*}[t]
	\centering
	\subfloat[Subject S1 (CN5, MNIST)]{\includegraphics[width=0.8\columnwidth]{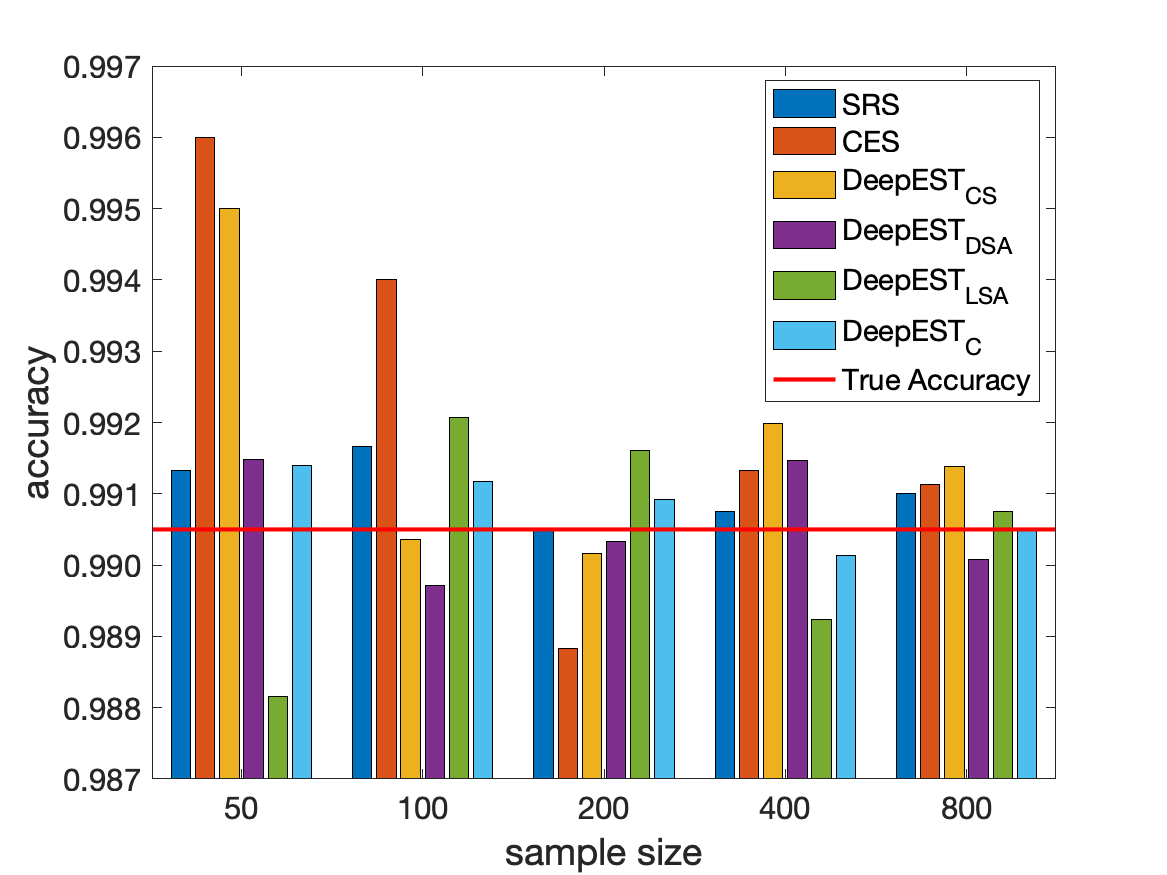}
	\label{fig:SMNISTmean}
	}
	\subfloat[Subject S5 (VGG16, CIFAR100)]{\includegraphics[width=0.8\columnwidth]{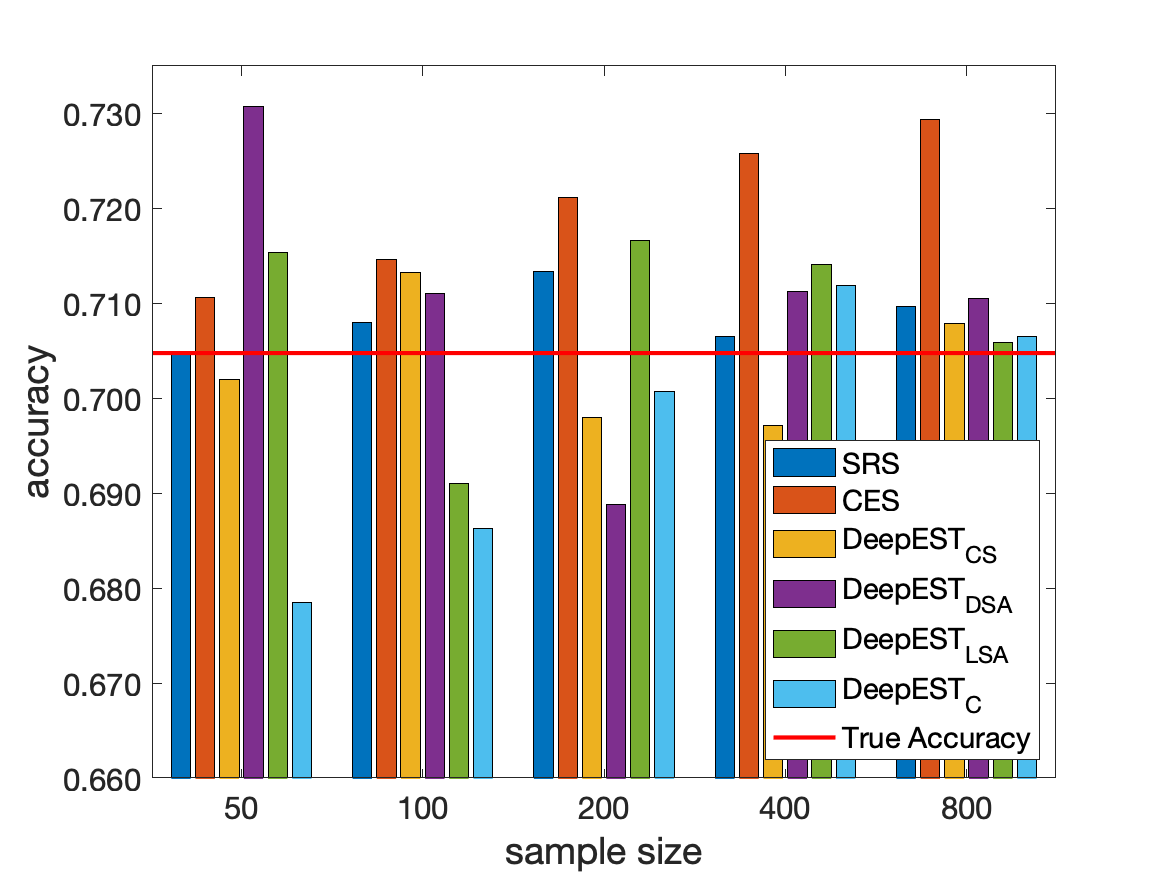}
	\label{fig:SCIFAR100mean}
	}
\vspace{-2pt}
	\caption{RQ2 (sensitivity to sample size): Accuracy for the most (a) and the least (b) accurate subjects}
	\label{fig:SMNISTSCIFAR100mean}
\end{figure*}

\begin{figure*}[htp]
\vspace{-15pt}
	\centering
	\subfloat[Subject S1 (CN5, MNIST)]{\includegraphics[width=0.9\columnwidth]{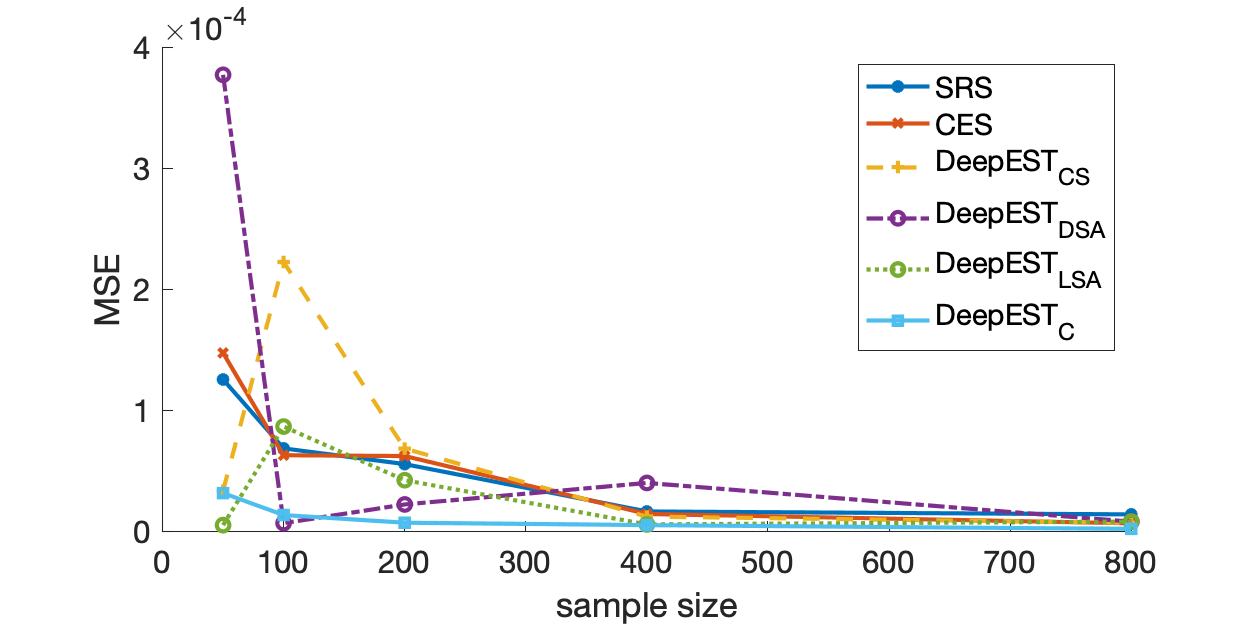}
		\label{fig:SMNISTMSE}}
	\subfloat[Subject S5 (VGG16, CIFAR100)]{\includegraphics[width=0.9\columnwidth]{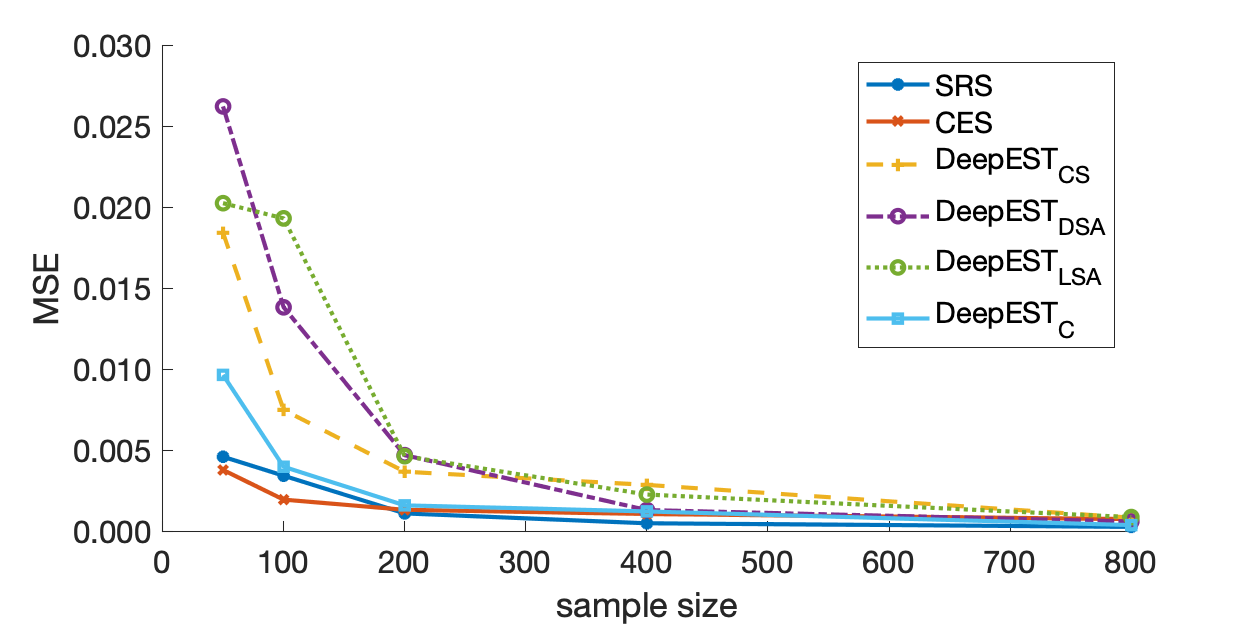}
		\label{}}
\vspace{-2pt}
	\caption{RQ2 (sensitivity to sample size): MSE for the most (a) and the least (b) accurate subjects}
	\label{fig:SMSE}
\end{figure*}

\begin{figure*}[htp]
\vspace{-15pt}
	\centering
	\subfloat[Subject S1 (CN5, MNIST)]{\includegraphics[width=0.9\columnwidth]{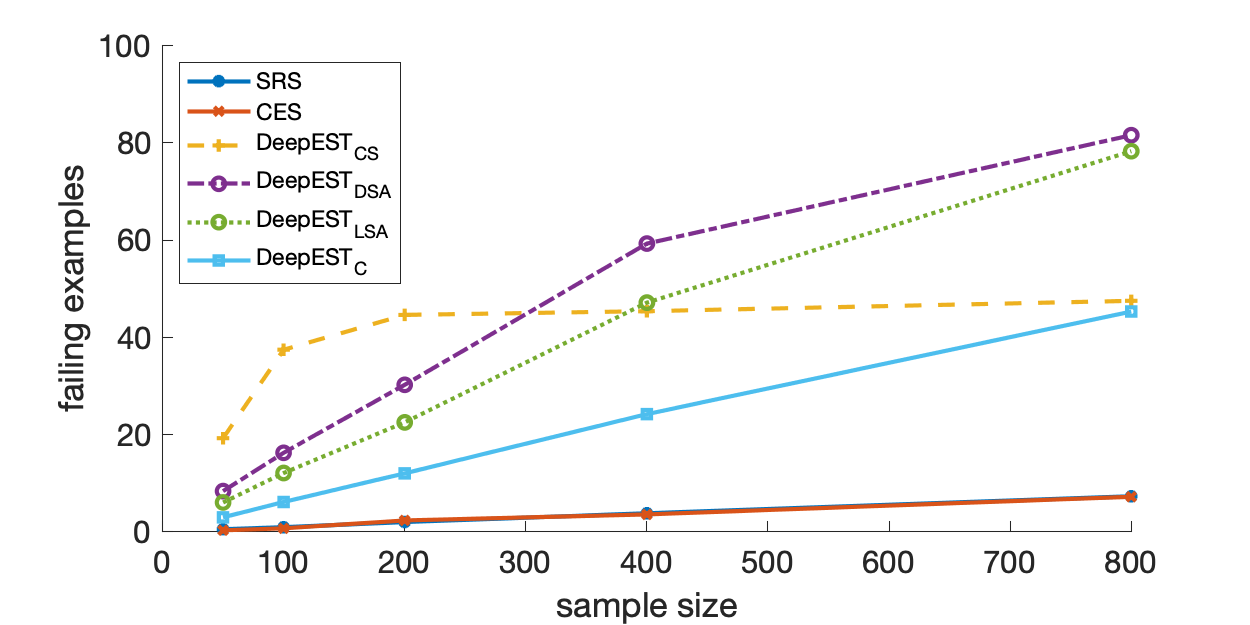}
		\label{fig:SMNISTFail}}
	\subfloat[Subject S5 (VGG16, CIFAR100)]{\includegraphics[width=0.9\columnwidth]{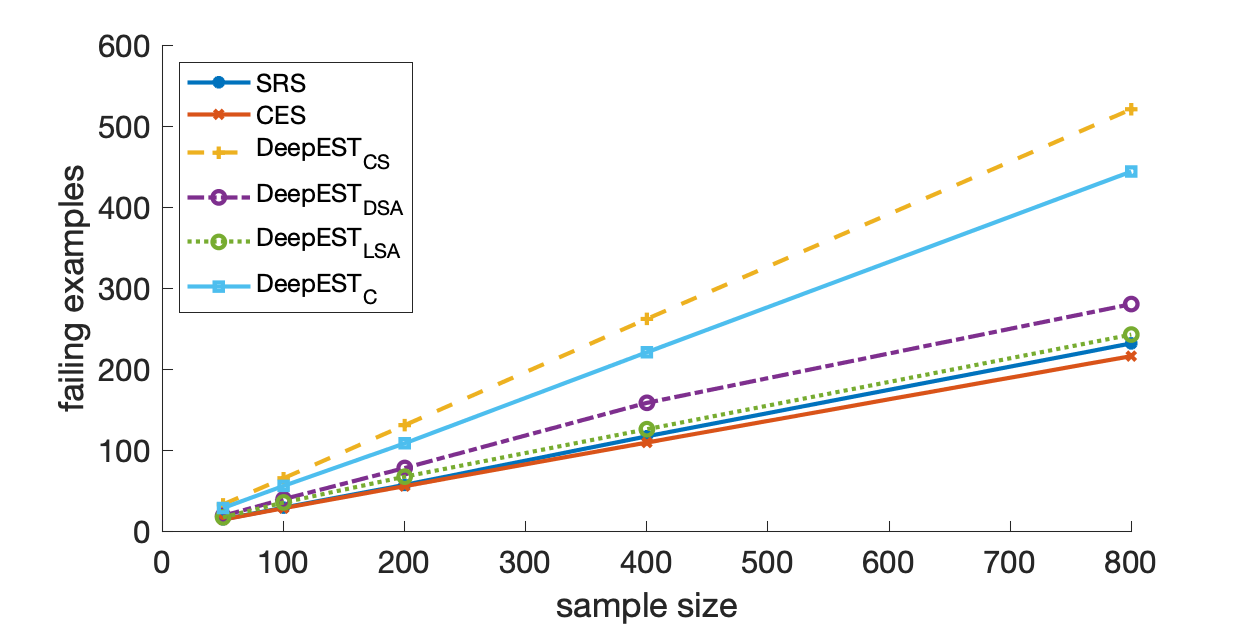}
		\label{}}
	\caption{RQ2 (sensitivity to sample size): Number of failing examples for the most (a) and the least (b) accurate subjects}
\vspace{-2pt}
	\label{fig:Sfe}
\end{figure*}

Table \ref{tabellone}(f) summarizes the number of wins (and losses).  CES never wins over other approaches, while \approach{}$_{C}$ wins against CES 3 out of 5 times.
SRS  never wins over \approach{}, while it wins \textit{vs} CES considering VGG16 on CIFAR100.
\approach{}$_C$ never looses and collects the highest number of wins (10), 
showing the best trade-off among accuracy of the estimation and number of detected failing examples. 
The choice of the \approach{} variant may be determined by which auxiliary information can be collected. 
We see that  \approach{}$_C$, exploiting a combination of two variables, has good and more stable results in terms of MSE than the other variants, at the expense of a slight decrease of detected failures. Single auxiliary variables are more sensitive to the specific dataset/model pair (e.g., confidence works well if the DNN is reliable), but expose more mispredictions. Confidence has the advantage of not requiring knowledge of the hidden layers and is easier to compute. When no information is available or easily computable, SRS could be a good low cost solution. %

\subsection{RQ2: sensitivity to sample size}

To answer this RQ, experiments are run with the sample sizes 50, 100, 200, 400, 800, considering the subject with the highest accuracy (S1), and the one with the lowest accuracy (S5), so as to analyze how \approach{} performs when there are very few and many failing examples in the dataset, respectively. %
Figure \ref{fig:SMNISTSCIFAR100mean} shows the mean values of the estimates' accuracy over repetitions. Figures \ref{fig:SMSE} and \ref{fig:Sfe} plot the MSE and the mean number of detected failures, respectively. Expectedly, increasing the sample size all techniques exhibit a decreasing trend in MSE and an increasing trend in failing examples.

For the subject with highest accuracy (S1), we observe the following. \approach{}$_{C}$ shows very good performance for MSE (Fig. \ref{fig:SMSE}(a), Fig. \ref{fig:MSE}), and it detects on average about six times the failures of SRS and CES (Table \ref{tab:numfp}).  
The advantage for MSE is more pronounced with the smallest sample sizes, which make \approach{}$_{C}$ particularly suited when %
the number of examples to select and label is very small. 
For larger sample sizes, the MSE is similar but the advantage of \approach{}$_{C}$ over SRS and CES is very pronounced for detected failures (Fig. \ref{fig:Sfe}(a)). 
\approach{}$_{CS}$ %
is the best among all techniques to detect failures for small sample sizes; for sizes $400$ and $800$, the best is \approach{}$_{DSA}$ (Fig. \ref{fig:Sfe}(a)).
Although the estimates are unbiased (hence, they tend to the true value), if we look at the mean estimates over the repetitions (Figure \ref{fig:SMNISTSCIFAR100mean}(a)), CES shows bad performance with up to $200$ test cases. SRS works well with low budget; its good results may be influenced by the very low number of failures: 18/30 repetitions show $0$ failures and $100\%$ accuracy.
It is interesting to note that in most cases CES and SRS overestimate the true accuracy -- an undesired property, especially for critical systems. This is related to the low number of failures detected, as discussed in Section \ref{Background}.

For the subject with lowest accuracy (S5), CES and SRS outperform \approach$_C$ for small sample sizes ($50$ and $100$); from $200$ to $800$, the estimation by CES starts diverging, while SRS and \approach{} keep good performance. The tendency to overestimate the accuracy by CES and SRS is confirmed. Performance in failing examples detection is always clearly in favour of all \approach{} variants. 
Performance in estimation accuracy is almost specular to what observed with the most accurate model.  A reason is that \approach{} is a sampling techniques particularly suited for rare populations, which is not the case of S5. 
As for the ability to detect failing examples, \textit{confidence} is the best auxiliary variable for \approach{} for subject S5: it presents the best values in all configurations. %

\subsection{RQ3: dataset influence}
We have seen in the experiments for RQ1 and RQ2 that no single auxiliary variable performs best in all situations. For instance, we can consider the confidence a good auxiliary information for subject S1, and a bad choice for S5. This may depend on several factors: assuming a perfect training, the confidence could be affected by a bias in the training set; or, with a perfect training set, a wrong training phase (e.g., due to overfitting) could generate mispredictions with high confidence. In some cases the operational dataset could contain examples very similar (i.e., small distance) to those in the training set but with a different label, affecting the discriminative power of the DSA and LSA metrics. 
To analyze how the three datasets influence the ability of auxiliary variables to discriminate failing examples (impacting the performance observed in RQ1-RQ2), we consider the subjects S1 and S5, as for 
 RQ2, plus the VGG16 DNN for CIFAR10.  %

\begin{figure*}[tp]
	\centering
	\subfloat[Confidence]{\includegraphics[width=0.3\textwidth]{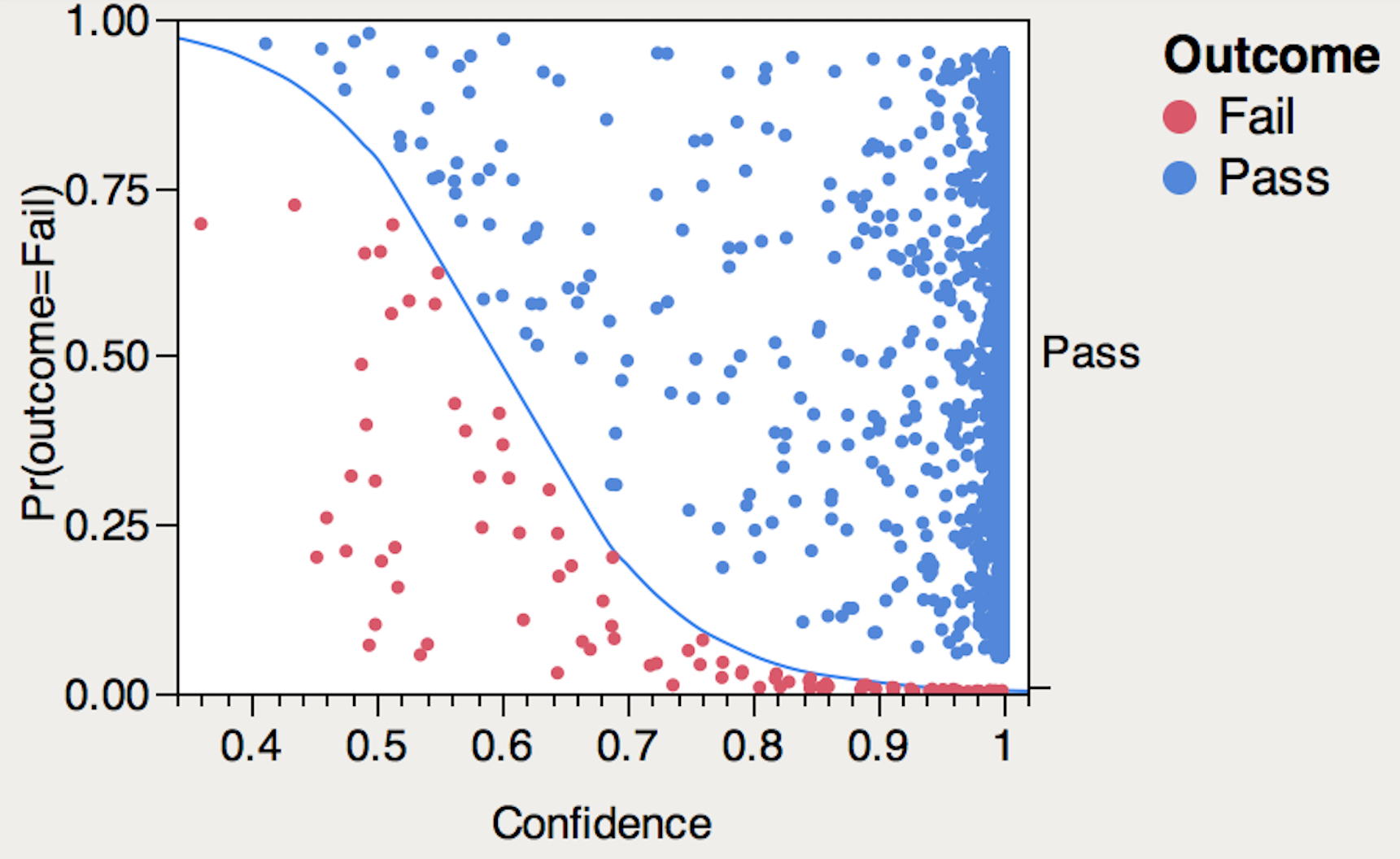}
		\label{}}
	\subfloat[DSA]{\includegraphics[width=0.3\textwidth]{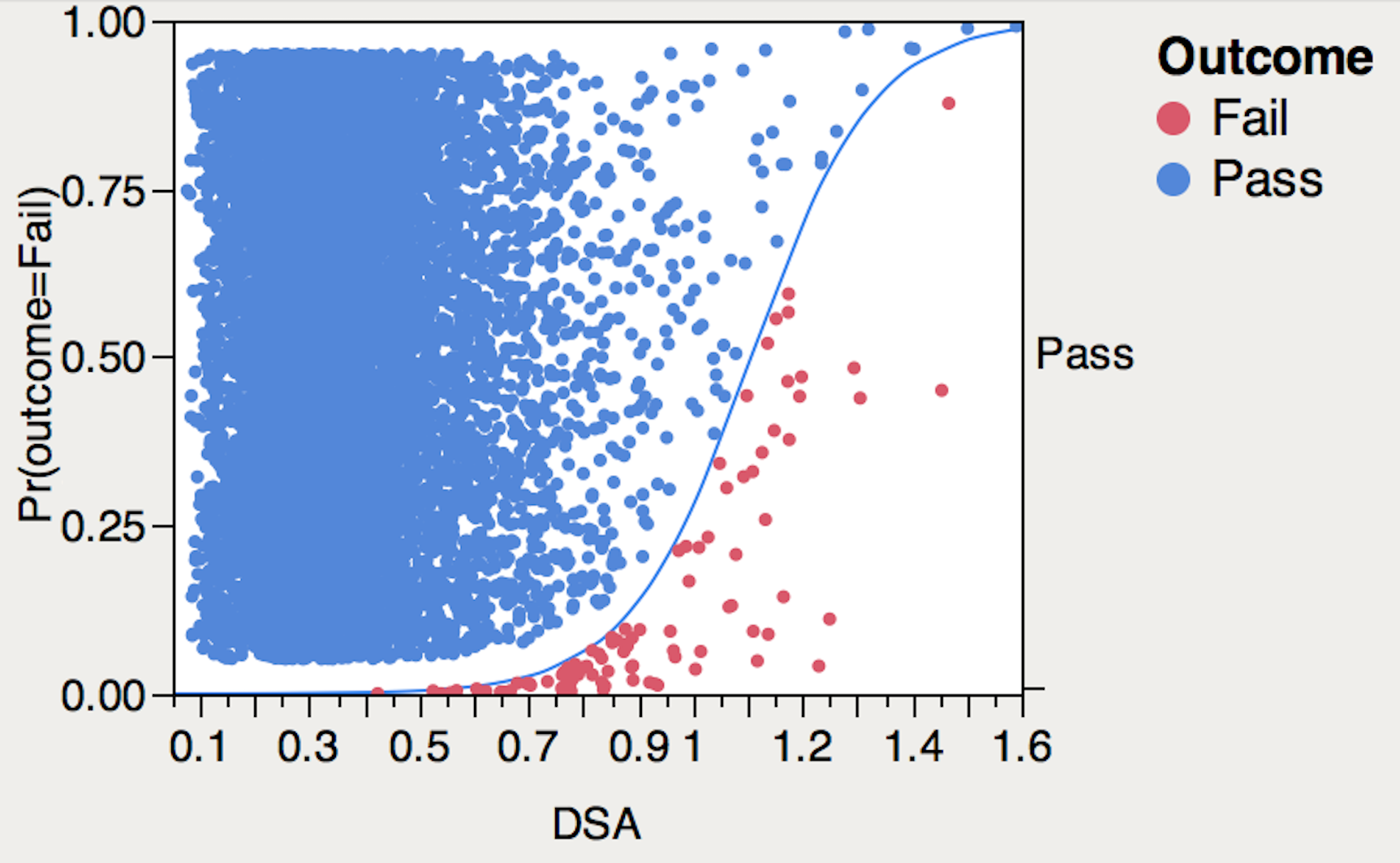}
		\label{}}
	\subfloat[LSA]{\includegraphics[width=0.3\textwidth]{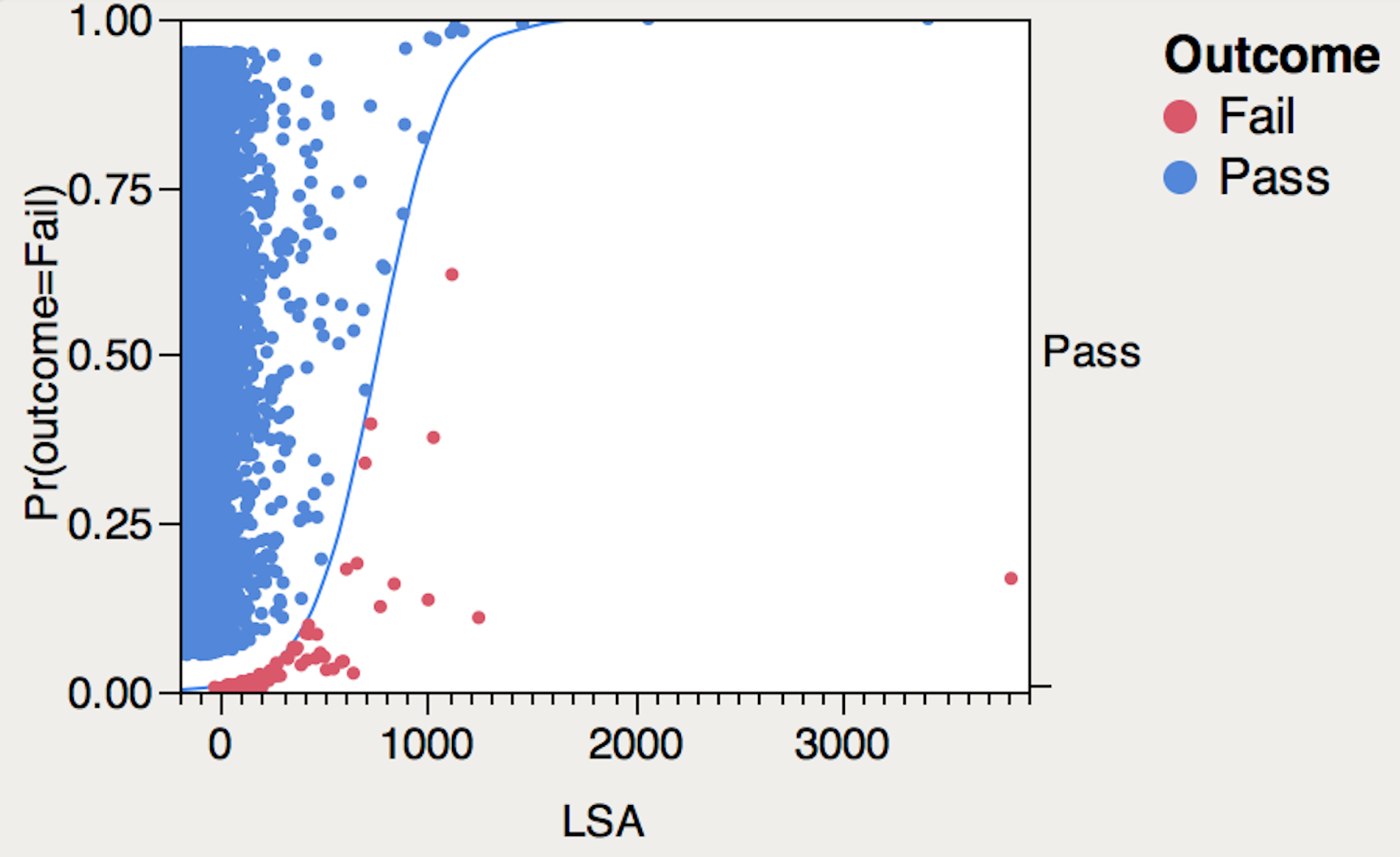}
		\label{}}
	\caption{RQ3 (dataset influence): MNIST samples distribution}
	\label{Msd}
\end{figure*}

\begin{figure*}[htp]
\vspace{-12pt}
	\centering
	\subfloat[Confidence]{\includegraphics[width=0.3\textwidth]{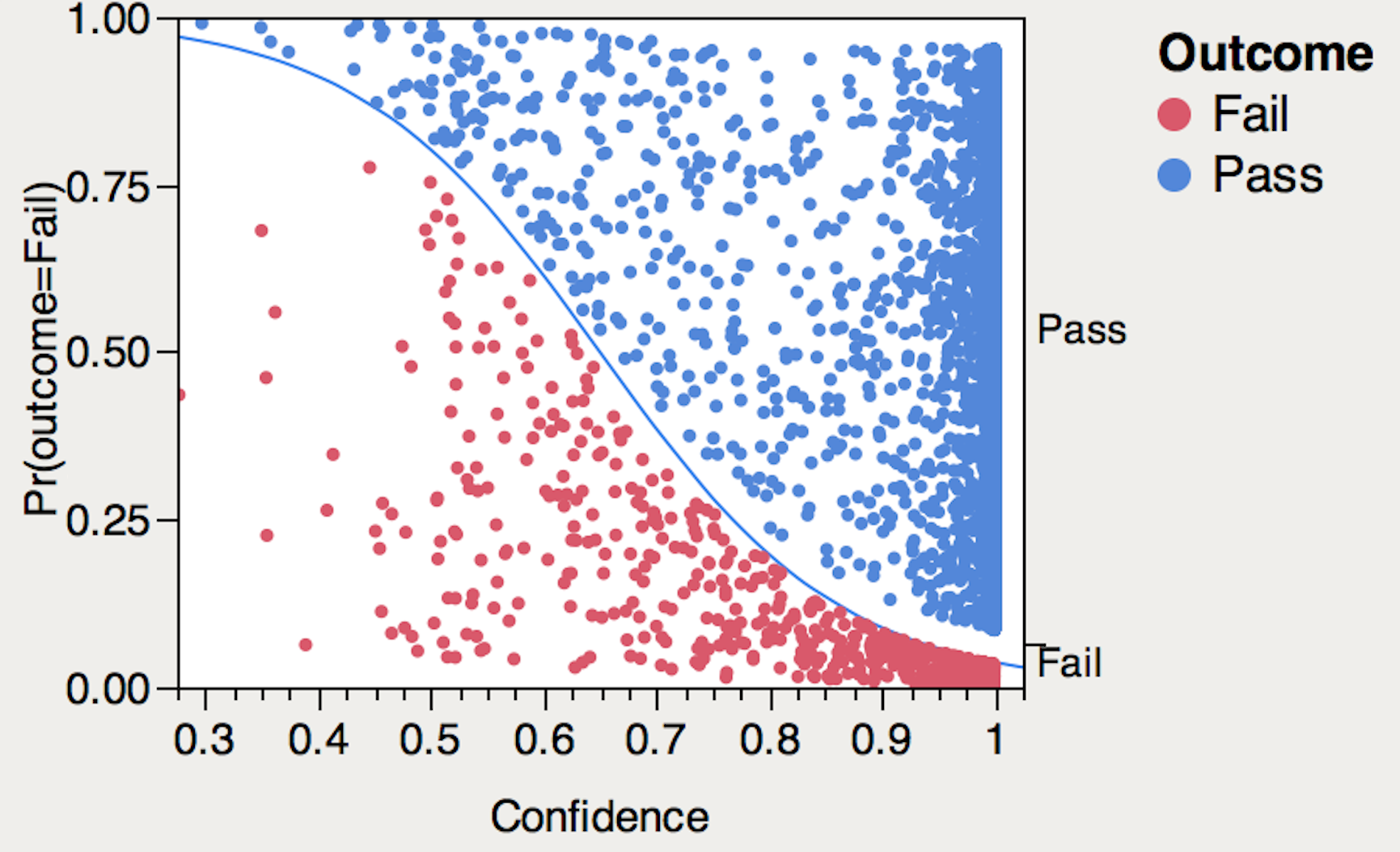}
		\label{}}
	\subfloat[DSA]{\includegraphics[width=0.3\textwidth]{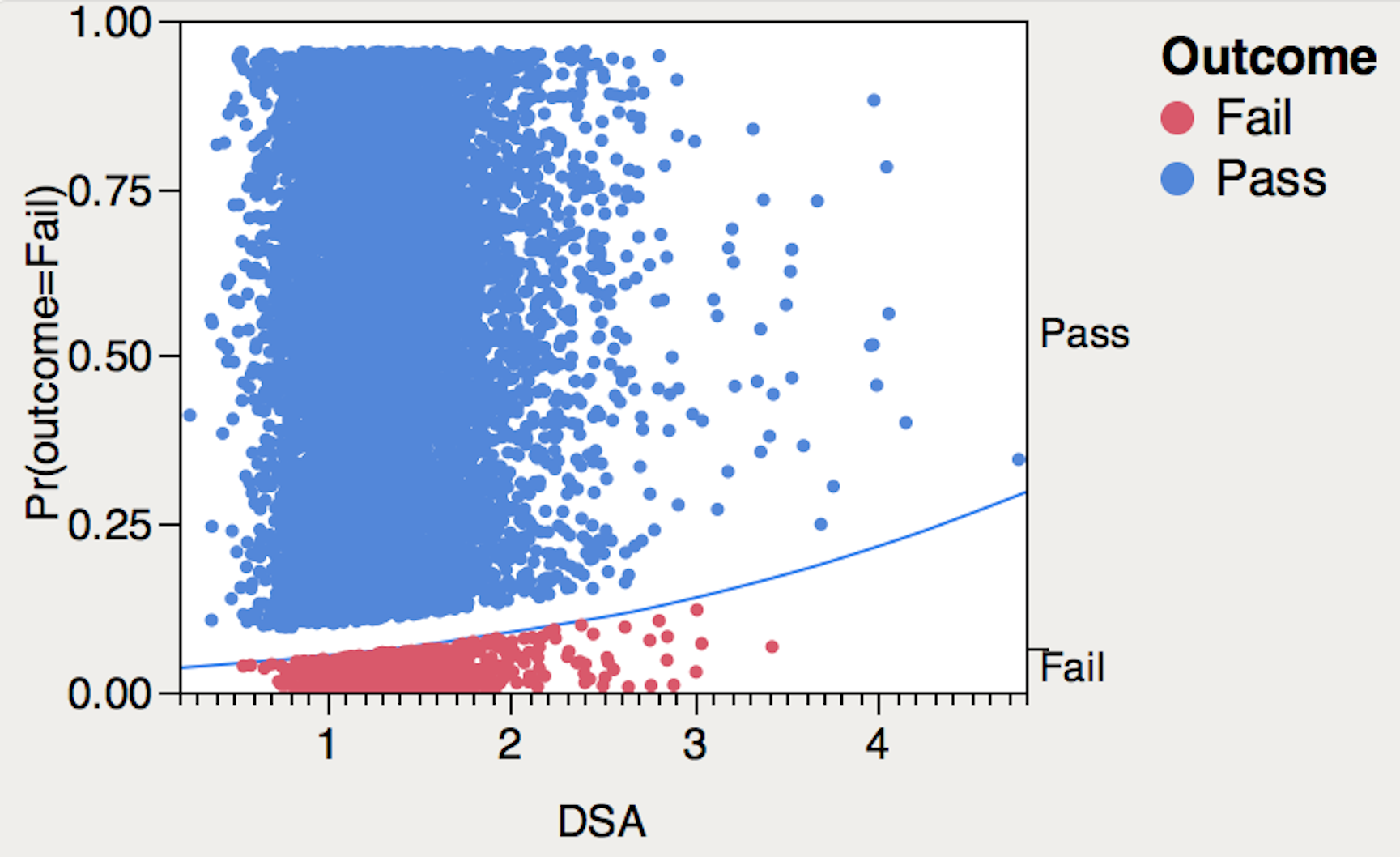}
		\label{}}
	\subfloat[LSA]{\includegraphics[width=0.3\textwidth]{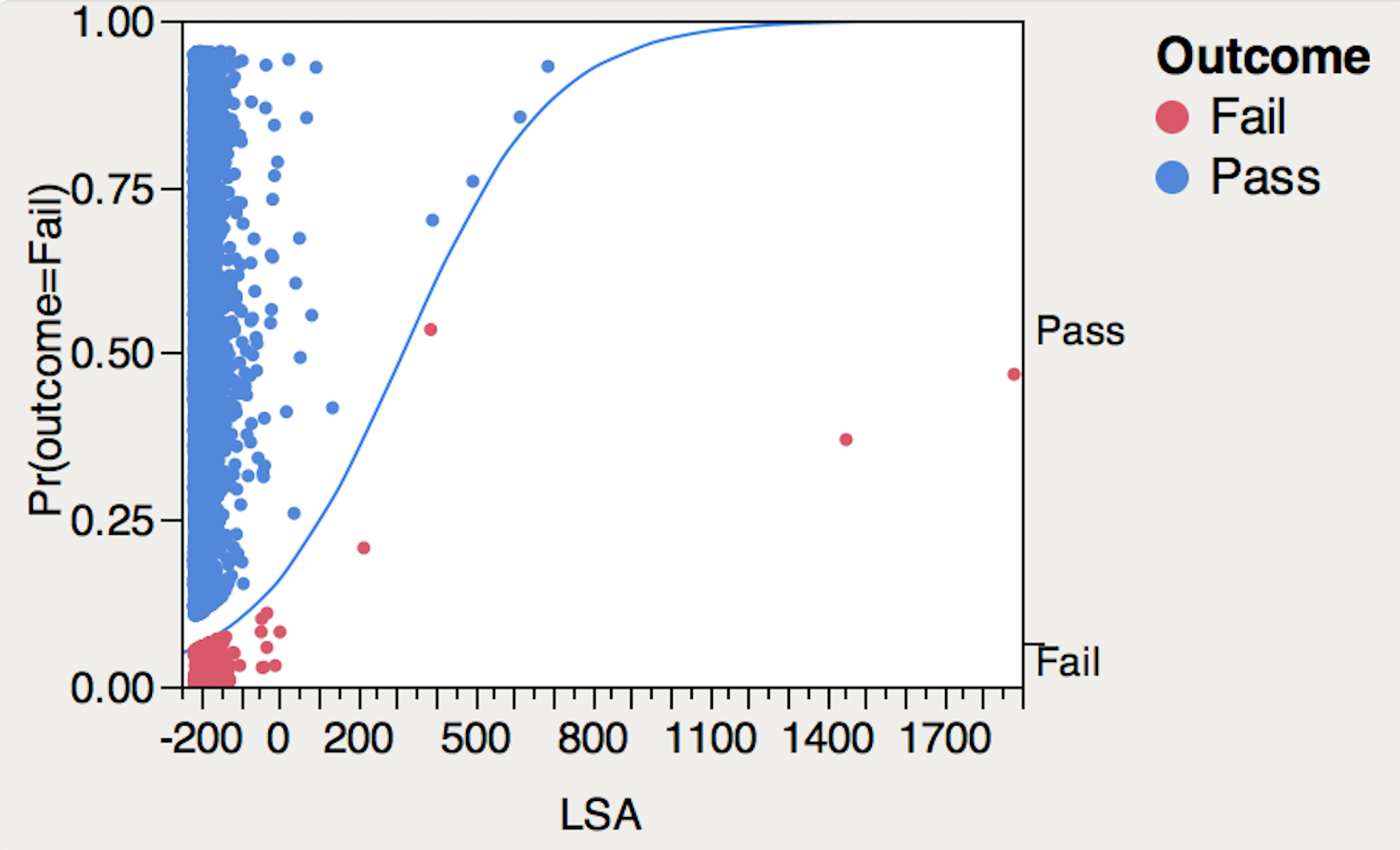}
		\label{}}
	\caption{RQ3 (dataset influence): CIFAR10 samples distribution}
	\label{C10sd}
\end{figure*}

\begin{figure*}[htp]
\vspace{-12pt}
	\centering
	\subfloat[Confidence]{\includegraphics[width=0.3\textwidth]{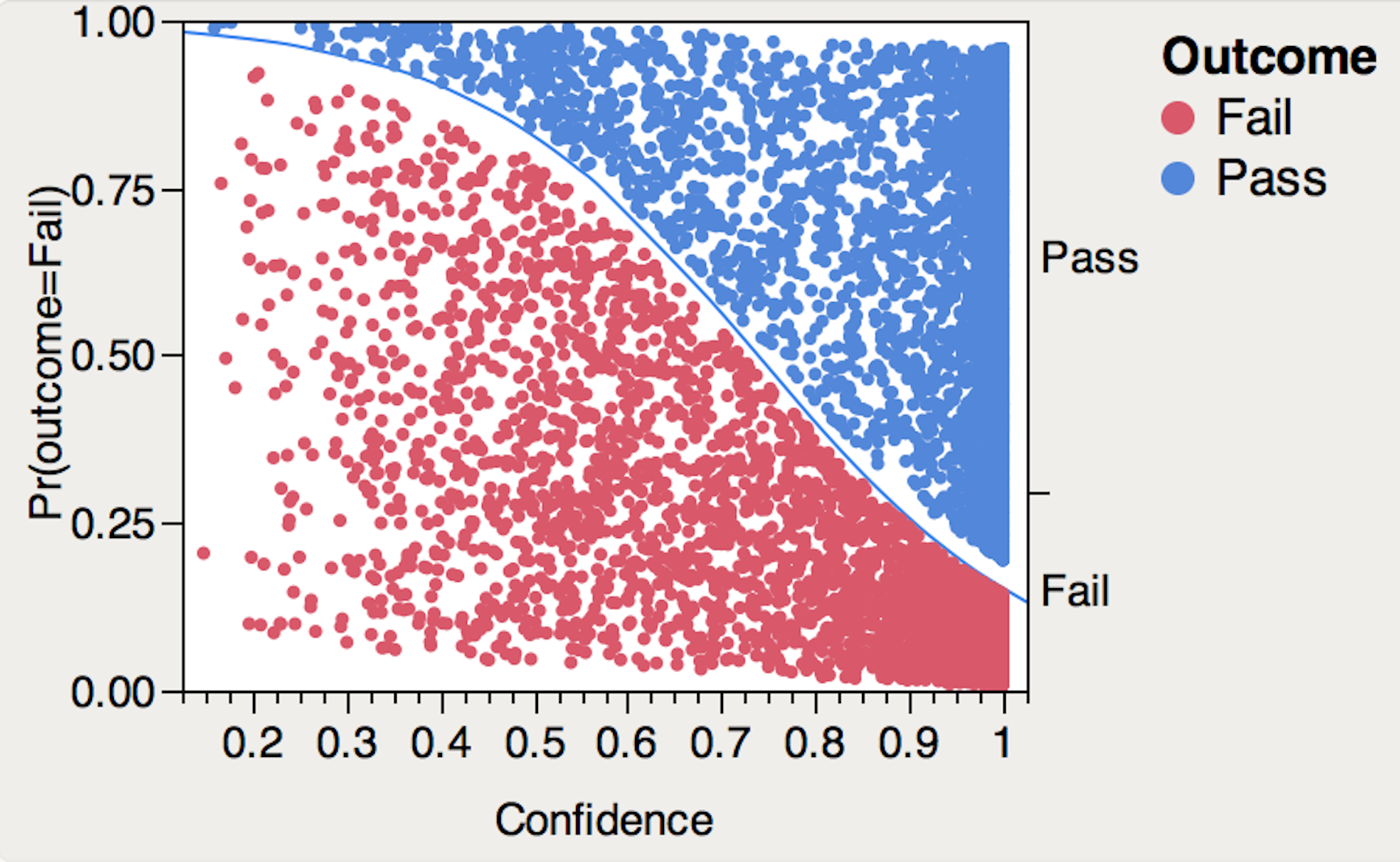}
		\label{}}
	\subfloat[DSA]{\includegraphics[width=0.3\textwidth]{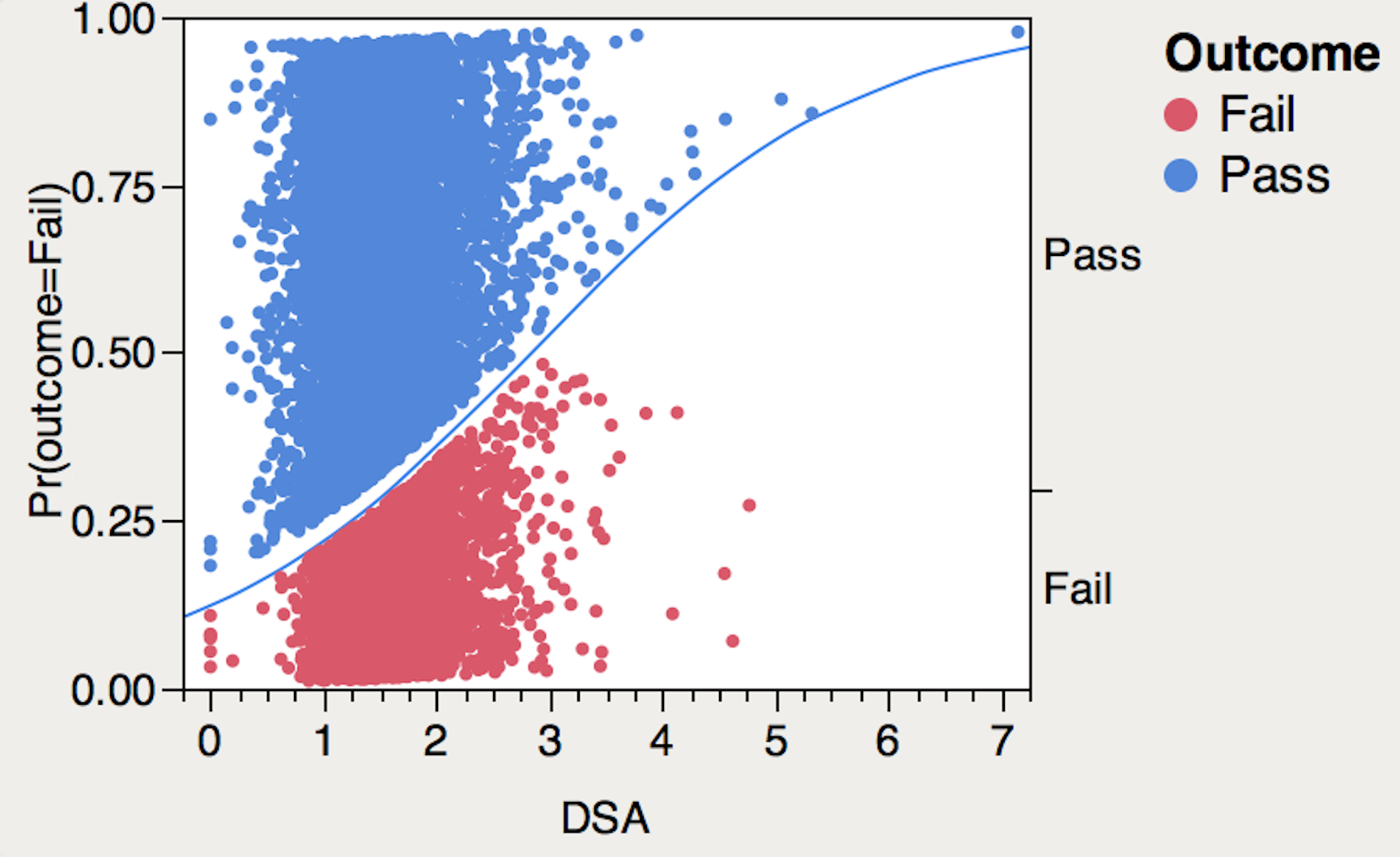}
		\label{}}
	\subfloat[LSA]{\includegraphics[width=0.3\textwidth]{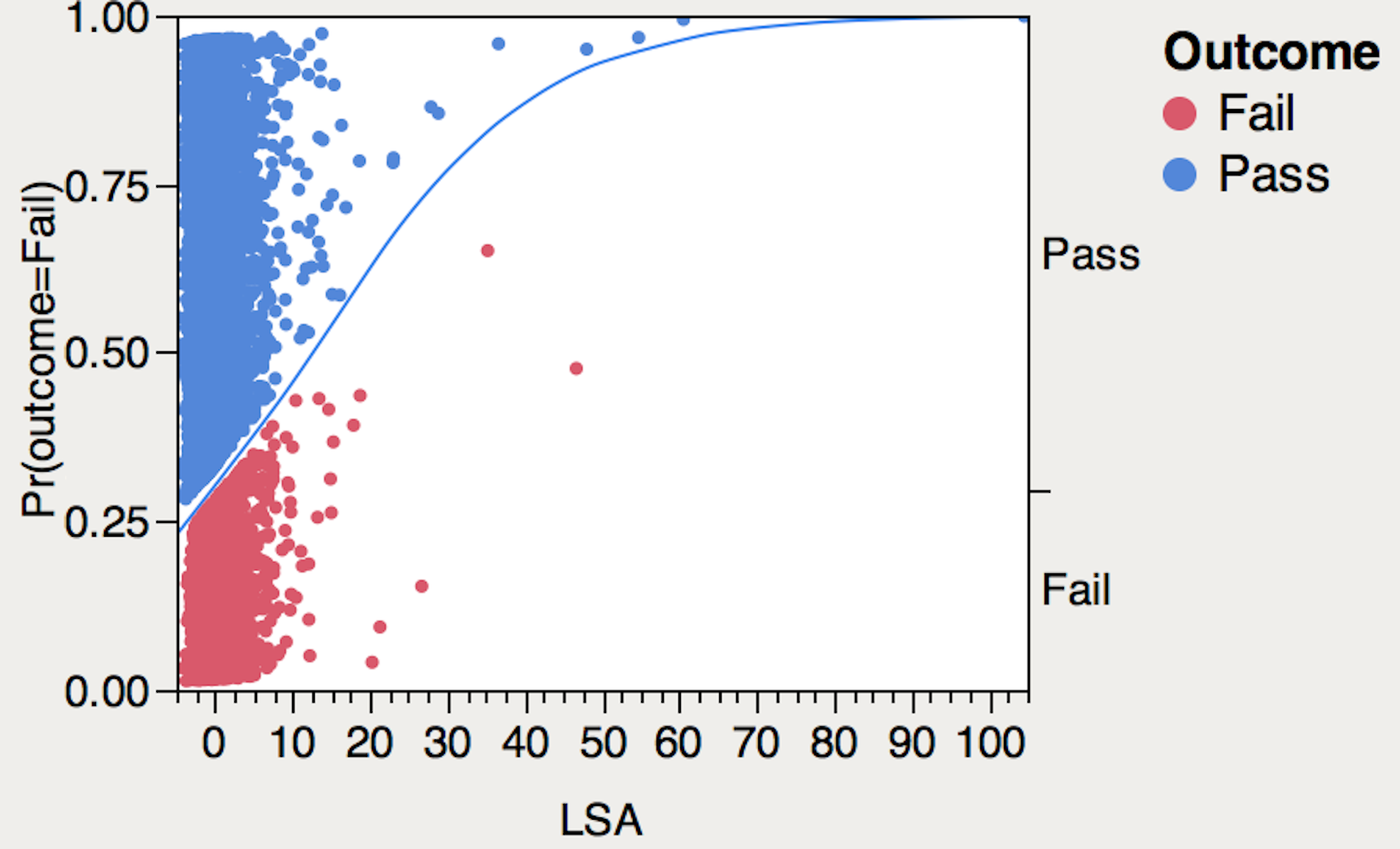}
		\label{}}
	\caption{RQ3 (dataset influence): CIFAR100 samples distribution}
\vspace{-6pt}
	\label{Csd}
\end{figure*}

Figures \ref{Msd}, \ref{C10sd}, and \ref{Csd} show the logistic regression for the three datasets. The curves fit the probabilities for the outcome fail/pass to the three predictors: confidence, DSA, and LSA.  
Consider MNIST, for which CN5 reaches the highest accuracy among all subjects; the probability for a test to fail is very low for values of confidence between 0.7 and 1 (Figure \ref{Msd}). 
This is clearly not the case for CIFAR10 (Figure \ref{C10sd}) and, especially, CIFAR 100 (Figure \ref{Csd}). In the latter, there is a high chance of misprediction even with high values of confidence; this could be due to a high skew between training and test data.

The discriminating power of DSA and LSA is clearly greater for MNIST, as the high slope of the S-shaped curves in Figures \ref{Msd}(b)-(c) suggests, compared to corresponding ones in Figures \ref{C10sd} and \ref{Csd}. In this case, it actually happens that 
the farthest examples have highest probability to be related to a failure, with a sharp increase after DSA$\approx$$0.9$ and LSA$\approx$$300$. 
This means that if in operation there are (a lot of) examples far from what observed in training, re-training with a more representative dataset can be useful to improve the accuracy.
This behaviour is not observed for CIFAR10 and CIFAR100, and DSA and LSA do not seem to be effective in dividing the two sets of examples. In CIFAR10, the DSA line is more horizontal, meaning that the DSA value does not reflect well the failure probability. 
The scarcely discriminative power of the auxiliary variables in CIFAR10 and CIFAR100 partially explains the smaller gain of \approach{} over CES and SRS (especially in terms of MSE); nevertheless,   
 its adaptivity allows spotting many failing examples even in these conditions. 

Finally, it is interesting to highlight the performance of \approach{}$_{CS}$ (based on confidence) on MNIST in RQ2. 
The saturation in its failure detection ability (Figure \ref{fig:Sfe}(a)) can be explained observing that, after a number of tests able to spot failures looking at low-confidence examples, the few remaining ones with high confidence are selected with low probability; the sharper discrimination made by DSA and LSA determines a high detection ability even when few failing examples remain.  
In summary, whenever a tester has good belief/evidence about the appropriateness of one of the above auxiliary variables, it is a good choice to select the specific \approach{} variant; if not, the combined variant \approach{}$_C$ has shown to give the best trade-offs in all five experimented cases. 

\subsection{Threats to validity}
A threat to the internal validity comes from the selection of the experimental subjects. To favor the repeatability of the experiments under different possibly influencing factors, we have used publicly available networks and pre-trained models, to avoid incorrect implementation. The configuration of parameter $r$ in \approach{} and a different setting of thresholds may also affect the results (in terms of efficiency of the estimator), hence a fine-tuning is suggested before applying the method to other dataset-DNNs. Although the  code developed was carefully inspected, a common threat
  is the correctness of the scripts to collect data and compute the results. 

The choice of the sample size influences the effectiveness too. We ran a sensitivity analysis to show that \approach{} is still effective (compared to both CES and SRS) under five (from 50 to 800) values of the sample size, but different values could yield different results.   
Threats to external validity depend on both the number of models and datasets considered for experimentation. We strived to control this threat considering different widely used DNN and datasets. Although the results may change with different subjects, the diversity and significance of the chosen subjects give confidence to the general considerations. Replicability of the experiments on other subjects is to further mitigate this threat.

\section{Related work}
\label{Related}
\makeatletter{}

Testing of DNN 
is a hot research topic. %
Zhang et al. \cite{Zhang20} present a survey on Machine Learning testing, identifying three main families of techniques: mutation testing \cite{Ma18DeepMutation}, metamorphic testing \cite{Xie11,Zhang18}, %
and cross referencing \cite{Srisak18, Pei19}. 
The former two are for test generation: they generate adversarial examples causing mispredictions. %
Cross-referencing can highlight the most interesting test cases when different implementations of a system disagree. 
Most of these techniques are meant for fault detection, rather than for accuracy estimate. 

Testing DNN for accuracy is the focus of Li et \textit{al.} \cite{Li19}, who presented CES -- which \approach{} has been compared against  - as the first 
approach using operational testing to this aim. \approach{} differs from CES in several aspects, the key ones being the sampling algorithm and the used auxiliary variables, which are conceived to improve failing examples detection while preserving the accuracy estimation power. 

Operational testing, where tests are derived according to the operational profile, is an established practice to estimate software reliability \cite{SRET1996}. 
It was the core technique of Cleanroom software engineering \cite{mills1, Currit1986, mills3, mills4} 
and of the Software Reliability Engineering Test process 
\cite{SRET1996}. %
Cai et \textit{al.} developed \textit{Adaptive Testing}, still based on the operational profile, but foreseeing adaptation in assigning  tests to partitions \cite{CaiTSE, CaiJSS, cai7, cai5}. %
Recently, Pietrantuno et \textit{al.} stressed the use of unequal probability sampling to improve the estimation efficiency \cite{QRS2018}, to this aim formalizing several sampling schemes \cite{ISSRE2016}. 
Our proposal goes along this direction: we do not sample tests representative of the operational dataset, but we ``alter'' the selection and counterbalance the uneven selection in the estimator. Unequal probability, without-replacement and adaptive sampling are the key concepts we borrowed for operational testing of DNNs.

\section{Conclusion}
\label{Conclusion}
\makeatletter{}
Testing the accuracy of a DNN with operational data aims at precise estimates with small test suites, due to the cost for manually labelling the selected test cases. This effort motivates to pursue also the goal of exposing many mispredictions with the same test suite, so as to improve the DNN after testing. 
With these two concurrent goals, we presented \approach{}, a technique to select failing tests from an operational dataset, while ultimately yielding faithful estimates of the accuracy of a DNN under test. 

We evaluated experimentally four variants of \approach{}, based on various types of auxiliary information that its adaptive sampling strategy can leverage. 
The results with four DNN models and three popular datasets show that all \approach{} variants provide accurate estimates, compared to existing sampling-based DNN testing techniques, while generally much outperforming them in exposing mispredictions. Practitioners may choose the appropriate variant, depending on the characteristics of their operational dataset and on which auxiliary information is available or they can collect.

\bibliographystyle{unsrt}
\bibliography{ICSE2021}

\begin{thebibliography}{10}

\bibitem{Obermeyer2016}
Ziad Obermeyer and Ezekiel~J. Emanuel.
\newblock Predicting the future --- big data, machine learning, and clinical
  medicine.
\newblock {\em New England Journal of Medicine}, 375(13):1216--1219, 2016.
\newblock PMID: 27682033.

\bibitem{Bojarski}
Mariusz~Bojarski \textit{et al.}
\newblock {End to End Learning for Self-Driving Cars}.
\newblock arXiv:1604.07316, 2016.

\bibitem{Amir2018}
Amir Efrati.
\newblock Uber finds deadly accident likely caused by software set to ignore
  objects on road.
\newblock {\em The Information}, May 7, 2018.

\bibitem{Stewart2018}
Jack Stewart.
\newblock Tesla's autopilot was involved in another deadly car crash.
\newblock [Online]. Available:
  https://www.wired.com/story/tesla-autopilot-self-driving-crash-california/,
  2018.

\bibitem{DeepTest}
Yuchi Tian, Kexin Pei, Suman Jana, and Baishakhi Ray.
\newblock {DeepTest}: Automated testing of deep-neural-network-driven
  autonomous cars.
\newblock In {\em 40th International Conference on Software Engineering}, ICSE,
  pages 303--314. ACM, 2018.

\bibitem{Taigman2014}
Yaniv {Taigman}, Ming {Yang}, {Marc'Aurelio} {Ranzato}, and Lior {Wolf}.
\newblock {DeepFace}: Closing the gap to human-level performance in face
  verification.
\newblock In {\em IEEE Conference on Computer Vision and Pattern Recognition},
  CVPR, pages 1701--1708. IEEE, 2014.

\bibitem{Pei19}
Kexin Pei, Yinzhi Cao, Junfeng Yang, and Suman Jana.
\newblock Deepxplore: Automated whitebox testing of deep learning systems.
\newblock {\em Communications of the ACM}, 62(11):137--145, 2019.

\bibitem{Ma18}
Lei Ma, Felix Juefei-Xu, Fuyuan Zhang, Jiyuan Sun, Minhui Xue, Bo~Li, Chunyang
  Chen, Ting Su, Li~Li, Yang Liu, Jianjun Zhao, and Yadong Wang.
\newblock {DeepGauge: Multi-Granularity Testing Criteria for Deep Learning
  Systems}.
\newblock In {\em 33rd ACM/IEEE International Conference on Automated Software
  Engineering}, ASE, pages 120--131. ACM, 2018.

\bibitem{Zhang18}
Mengshi {Zhang}, Yuqun {Zhang}, Lingming {Zhang}, Cong {Liu}, and Sarfraz
  {Khurshid}.
\newblock {DeepRoad: GAN-Based Metamorphic Testing and Input Validation
  Framework for Autonomous Driving Systems}.
\newblock In {\em 33rd ACM/IEEE International Conference on Automated Software
  Engineering}, ASE, pages 132--142. ACM, 2018.

\bibitem{Ma2018combinatorial}
Lei Ma, Fuyuan Zhang, Minhui Xue, Bo~Li, Yang Liu, Jianjun Zhao, and Yadong
  Wang.
\newblock Combinatorial testing for deep learning systems.
\newblock arxiv.org/abs/1806.07723, 2018.

\bibitem{Odena18}
Augustus Odena and Ian Goodfellow.
\newblock {TensorFuzz}: Debugging neural networks with coverage-guided fuzzing.
\newblock In {\em 36th International Conference on Machine Learning}, volume~97
  of {\em Proc. of Machine Learning Research}, 2019.

\bibitem{Li19Structural}
Zenan Li, Xiaoxing Ma, Chang Xu, and Chun Cao.
\newblock Structural coverage criteria for neural networks could be misleading.
\newblock In {\em 41st International Conference on Software Engineering: New
  Ideas and Emerging Results}, ICSE-NIER, pages 89--92. IEEE, 2019.

\bibitem{Wu2019}
Weibin {Wu}, Hui {Xu}, Sanqiang {Zhong}, Michael~R. {Lyu}, and Irwin {King}.
\newblock {Deep Validation: Toward Detecting Real-World Corner Cases for Deep
  Neural Networks}.
\newblock In {\em 49th Annual IEEE/IFIP International Conference on Dependable
  Systems and Networks}, DSN, pages 125--137. IEEE, 2019.

\bibitem{Kim19}
Jinhan Kim, Robert Feldt, and Shin Yoo.
\newblock {Guiding Deep Learning System Testing Using Surprise Adequacy}.
\newblock In {\em 41st International Conference on Software Engineering}, ICSE,
  pages 1039--1049. IEEE, 2019.

\bibitem{Frankl1998}
Phyllis~G. Frankl, Richard~G. Hamlet, Bev Littlewood, and Lorenzo Strigini.
\newblock {Evaluating testing methods by delivered reliability}.
\newblock {\em IEEE Transactions on Software Engineering}, 24(8):586--601,
  1998.

\bibitem{Lyu1996}
Michael~R. Lyu, editor.
\newblock {\em Handbook of Software Reliability Engineering}.
\newblock McGraw-Hill, Inc., Hightstown, NJ, USA, 1996.

\bibitem{Li19}
Zenan Li, Xiaoxing Ma, Chang Xu, Chun Cao, Jingwei Xu, and Jian L\"{u}.
\newblock {Boosting Operational DNN Testing Efficiency through Conditioning}.
\newblock In {\em Proc. of the 2019 27th ACM Joint Meeting on European Software
  Engineering Conference and Symposium on the Foundations of Software
  Engineering}, ESEC/FSE, pages 499--509. ACM, 2019.

\bibitem{TSE2016}
Domenico Cotroneo, Roberto Pietrantuono, and Stefano Russo.
\newblock {RELAI testing: a technique to assess and improve software
  reliability}.
\newblock {\em IEEE Transactions on Software Engineering}, 42(5):452--475,
  2016.

\bibitem{ISSRE2016}
Roberto Pietrantuono and Stefano Russo.
\newblock On adaptive sampling-based testing for software reliability
  assessment.
\newblock In {\em 27th International Symposium on Software Reliability
  Engineering}, ISSRE, pages 1--11. IEEE, 2016.

\bibitem{BOOK1}
Sharon~L. Lohr.
\newblock {\em Sampling: Design and Analysis}.
\newblock Duxbury Press, 2009.

\bibitem{BOOKThompson}
Steven~K. Thompson.
\newblock {\em Sampling, Third Edition}.
\newblock John Wiley \& Sons, Inc., 2012.

\bibitem{Thompson}
Daniel~G. Horvitz and Donovan~J. Thompson.
\newblock A generalization of sampling without replacement from a finite
  universe.
\newblock {\em Journal of the American Statistical Association}, 47(260):pp.
  663--685, 1952.

\bibitem{Hansen1943}
Morris~H. Hansen and William~N. Hurwitz.
\newblock {On the Theory of Sampling from Finite Populations}.
\newblock {\em The Annals of Mathematical Statistics}, 14(4):333--362, 1943.

\bibitem{Lecun98}
Yann {Lecun}, L\'eon {Bottou}, Yoshua {Bengio}, and Patrick {Haffner}.
\newblock Gradient-based learning applied to document recognition.
\newblock {\em Proceedings of the IEEE}, 86(11):2278--2324, 1998.

\bibitem{Krizhevsky09}
Alex Krizhevsky.
\newblock Learning multiple layers of features from tiny images.
\newblock Technical Report TR-2009, University of Toronto, 2009.

\bibitem{Zhang20}
Jie~M. Zhang, Mark Harman, Lei Ma, and Yang Liu.
\newblock Machine learning testing: Survey, landscapes and horizons.
\newblock {\em IEEE Transactions on Software Engineering}, pages 1--37, 2020.

\bibitem{Ma18DeepMutation}
Lei Ma, Fuyuan Zhang, Jiyuan Sun, Minhui Xue, Bo~Li, Felix Juefei-Xu, Chao Xie,
  Li~Li, Yang Liu, Jianjun Zhao, and et~al.
\newblock {DeepMutation: Mutation Testing of Deep Learning Systems}.
\newblock In {\em 29th International Symposium on Software Reliability
  Engineering}, ISSRE, pages 100--111. IEEE, 2018.

\bibitem{Xie11}
Xiaoyuan Xie, Joshua W.~K. Ho, Christian Murphy, Gail Kaiser, Baowen Xu, and
  Tsong~Yueh Chen.
\newblock Testing and validating machine learning classifiers by metamorphic
  testing.
\newblock {\em Journal of Systems and Software}, 84(4):544--558, 2011.

\bibitem{Srisak18}
Siwakorn Srisakaokul, Zhengkai Wu, Angello Astorga, Oreoluwa Alebiosu, and Tao
  Xie.
\newblock Multiple-implementation testing of supervised learning software.
\newblock In {\em AAAI Workshops}. Association for the Advancement of
  Artificial Intelligence, 2018.

\bibitem{SRET1996}
John~D. Musa.
\newblock Software reliability-engineered testing.
\newblock {\em Computer}, 29(11):61--68, 1996.

\bibitem{mills1}
Harlan~D. Mills, Michael Dyer, and Richard~C. Linger.
\newblock Cleanroom software engineering.
\newblock {\em IEEE Software}, 4(55):19--24, 1987.

\bibitem{Currit1986}
P.~Allen Currit, Michael Dyer, and Harlan~D. Mills.
\newblock Certifying the reliability of software.
\newblock {\em IEEE Transactions on Software Engineering}, SE-12(1):3--11,
  1986.

\bibitem{mills3}
Richard~H. Cobb and Harlan~D. Mills.
\newblock Engineering software under statistical quality control.
\newblock {\em IEEE Software}, 7(6):45--54, 1990.

\bibitem{mills4}
Richard~C. Linger and Harlan~D. Mills.
\newblock {A case study in cleanroom software engineering: the IBM COBOL
  Structuring Facility}.
\newblock In {\em 12th International Computer Software and Applications
  Conference}, COMPSAC, pages 10--17. IEEE, 1988.

\bibitem{CaiTSE}
Junpeng Lv, Bei-Bei Yin, and Kai-Yuan Cai.
\newblock {On the asymptotic behavior of adaptive testing strategy for software
  reliability assessment}.
\newblock {\em IEEE Transactions on Software Engineering}, 40(4):396--412,
  2014.

\bibitem{CaiJSS}
Junpeng Lv, Bei-Bei Yin, and Kai-Yuan Cai.
\newblock Estimating confidence interval of software reliability with adaptive
  testing strategy.
\newblock {\em Journal of Systems and Software}, 97:192--206, 2014.

\bibitem{cai7}
Kai-Yuan Cai, Chang-Hai. Jiang, Hai Hu, and Cheng-Gang Bai.
\newblock An experimental study of adaptive testing for software reliability
  assessment.
\newblock {\em Journal of Systems and Software}, 81(8):1406--1429, 2008.

\bibitem{cai5}
Kai-Yuan Cai, Yong-Chao Li, and Ke~Liu.
\newblock Optimal and adaptive testing for software reliability assessment.
\newblock {\em Information and Software Technology}, 46(15):989--1000, 2004.

\bibitem{QRS2018}
Roberto Pietrantuono and Stefano Russo.
\newblock Probabilistic sampling-based testing for accelerated reliability
  assessment.
\newblock In {\em International Conference on Software Quality, Reliability and
  Security}, QRS, pages 35--46. IEEE, 2018.

\end{thebibliography}

\end{document}